\documentclass[12pt]{article}
\usepackage{a4wide}
\usepackage{graphicx}

\usepackage{axodraw}

\newcommand{\dsp}{\displaystyle}
\newcommand{\be}{\begin{equation}}
\newcommand{\ee}{\end{equation}}
\newcommand{\ba}{\begin{eqnarray}}
\newcommand{\ea}{\end{eqnarray}}
\begin{document}
\begin{titlepage}
\begin{flushright}
LU TP 02-21\\
hep-ph/0205341\\
revised October 2002\\
\end{flushright}
\vfill
\begin{center}
{\Large\bf $K\to3\pi$ Decays in Chiral Perturbation Theory}\\
\vfill
{\bf Johan Bijnens, Pierre Dhonte and Fredrik Persson}\\[1cm]
{Department of Theoretical Physics, Lund University\\
S\"olvegatan 14A, S 22362 Lund, Sweden}
\end{center}
\vfill
\begin{abstract}
The CP conserving amplitudes for the
decays $K\to3\pi$ are calculated in Chiral Perturbation
Theory at the next-to-leading order. We present the expressions
in a compact form with single parameter functions only.
These expressions are then fitted to all available $K\to2\pi$ and
$K\to3\pi$ data to obtain a fit of the parameters that occur.
We compare with the work of Kambor, Missimer and Wyler.
\end{abstract}
\vfill
{\bf PACS numbers:} 13.20.Eb, 12.39.Fe, 14.40.Aq, 11.30.Rd

\end{titlepage}

\section{Introduction}
\label{introduction}

Chiral Perturbation Theory (ChPT) is the low-energy effective field theory
of the strong interactions. It was introduced in its modern form
by Weinberg, Gasser and Leutwyler \cite{Weinberg,GL1,GL2}. It has
had many successes and applications. A pedagogical introduction
can be found in \cite{chptlectures}.
The method has been used as well for nonleptonic weak decays. The
main work of extending ChPT to the nonleptonic weak interaction was 
done long ago by Kambor, Missimer and Wyler 
who worked out the general formalism \cite{KMW1} and applied it
to $K\to2\pi,3\pi$ decays \cite{KMW2}. These results were then used
to obtain directly relations between physical observables in
\cite{KDHMW}. Reviews of applications of ChPT to nonleptonic
weak interactions are \cite{chptweakreviews}.

Earlier work using current algebra methods or tree level Lagrangians
relevant for $K\to3\pi$ are \cite{Cronin,k3piold}
and references therein. Many new measurements in $K\to3\pi$ have
become available since the work of \cite{KMW2} so an update
of the fits of the parameters done there became necessary. The analytical
expressions of \cite{KMW2} were never published and have since been lost.
This forced us to reevaluate these decays and we present the analytical
results here using a simplified form derived using the arguments
first used for $\pi\pi$-scattering by Knecht et al.~\cite{knechtetal}.
We then perform a detailed comparison of the expressions with the data
and find a reasonable agreement. We point out some directions
for future work.

The next section 
defines our notation and gives a unique list of possible
terms at next-to-leading order in ChPT. Section \ref{kinematics}
gives the list of decays and the isospin constraints as well
as the simplified analytical form valid up till order $p^6$ in ChPT.
The ambiguities in this simplified form are discussed in detail
in App. \ref{App:ambiguity}.
This section also discusses the isospin constraints on the various amplitudes.
Our main result, the recalculation of the $K\to3\pi$ amplitudes
is discussed in Sect. \ref{analytical} where we present
the tree level result and work out the independent combinations
of the weak $p^4$ parameters that occur. 
The expressions at order $p^4$ are still rather cumbersome and are listed
in App.~\ref{App:results}.
Section \ref{dataplusfits} contains a short review of the available data
and presents in detail the various fits we have performed to the data.
Our conclusions are summarized in the final section.

\section{The ChPT Lagrangian}
\label{lagrangian}

The Lagrangian of ChPT in the strong sector was worked out at
next-to-leading order (NLO)
in \cite{GL1} and is given by
\be
{\cal L}_S = {\cal L}_{S2}+{\cal L}_{S4}\,,
\ee
with
\ba
\label{L2S}
{\cal L}_{S2} &=& \frac{F_0^2}{4} \{\langle D_\mu U^\dagger D^\mu U \rangle
+\langle \chi^\dagger U+\chi U^\dagger \rangle \}\, ,
\\
\label{LS4}
{\cal L}_{S4}&=&
L_1 \langle D_\mu U^\dagger D^\mu U \rangle^2
+L_2 \langle D_\mu U^\dagger D_\nu U \rangle
     \langle D^\mu U^\dagger D^\nu U \rangle \nonumber\\&&\hspace{-0.5cm}
+L_3 \langle D^\mu U^\dagger D_\mu U D^\nu U^\dagger D_\nu U\rangle
+L_4 \langle D^\mu U^\dagger D_\mu U \rangle \langle \chi^\dagger U
+\chi U^\dagger \rangle
\nonumber\\&&
+L_5 \langle D^\mu U^\dagger D_\mu U (\chi^\dagger U+U^\dagger \chi ) \rangle
+L_6 \langle \chi^\dagger U+\chi U^\dagger \rangle^2
\nonumber\\&&
+L_7 \langle \chi^\dagger U-\chi U^\dagger \rangle^2
+L_8 \langle \chi^\dagger U \chi^\dagger U 
+ \chi U^\dagger \chi U^\dagger \rangle
\nonumber\\&&
-i L_9 \langle F^R_{\mu\nu} D^\mu U D^\nu U^\dagger +
               F^L_{\mu\nu} D^\mu U^\dagger D^\nu U \rangle
\nonumber\\&&
+L_{10} \langle U^\dagger  F^R_{\mu\nu} U F^{L\mu\nu} \rangle\,.
\ea
$\langle A\rangle$ stands for the flavour trace of the matrix $A$,
and $F_0$ is the pion decay constant in the chiral limit.
The special unitary matrix $U$ contains the Goldstone boson fields
\be
U = \exp\left(\frac{i\sqrt{2}}{F_0}M\right)\,,\quad
M =\left(\begin{array}{ccc}
\frac{1}{\sqrt{2}}\pi^0+\frac{1}{\sqrt{6}}\eta & \pi^+ & K^+\\
\pi^- & \frac{-1}{\sqrt{2}}\pi^0+\frac{1}{\sqrt{6}}\eta & K^0\\
K^- & \overline{K^0} & \frac{-2}{\sqrt{6}}\eta
	 \end{array}\right)\,.
\ee
The formalism we use is the external field method of \cite{GL1}
with $s$, $p$, $l_\mu$ and $r_\mu$ matrix valued scalar, pseudo-scalar,
left-handed and right handed vector external fields respectively.
These show up in
\be
\chi = 2 B_0\left(s+ip\right)\,,
\ee
in the covariant derivative 
\be
\label{covariant}
D_\mu U = \partial_\mu U -i r_\mu U + i U l_\mu\,,
\ee
and in the field strength tensor
\be
F_{\mu\nu}^{L(R)} = \partial_\mu l(r)_\nu -\partial_\nu l(r)_\mu -i 
\left[ l(r)_\mu , l(r)_\nu \right]\,.
\ee
For our purpose it suffices to set
\be
s = 
\left(\begin{array}{ccc}m_u &  & \\ & m_d & \\ & & m_s\end{array}\right)\,,
%\nonumber\\
\quad p = 0\,,
\quad
l_\mu = r_\mu = 0\,.
\ee
We also define the matrices $u$,  $u_\mu$ and $\chi_{\pm}$,
\be
u_\mu = i u^\dagger D_\mu U u^\dagger = u_\mu^\dagger\,,\quad u^2 = U,
\quad \chi_{\pm} = u^\dagger \chi u^\dagger \pm u \chi^\dagger u\,.
\ee

The ChPT Lagrangian for the weak nonleptonic interactions at lowest order
dates back to current algebra days \cite{Cronin} and is
\be
\label{LW2}
{\cal L}_{W2} = C \, F_0^4 \, 
\Bigg[ G_8 \langle \Delta_{32} u_\mu u^\mu \rangle +
G_8' \langle\Delta_{32} \chi_+ \rangle  +
G_{27} t^{ij,kl} \, \langle \Delta_{ij } u_\mu \rangle
\langle\Delta_{kl} u^\mu \rangle \Bigg]
+ \mbox{ h.c.}\nonumber
\ee
The tensor $t^{ij,kl}$ has as nonzero components
\ba
\label{deft}
t^{21,13} =
t^{13,21} = \frac{1}{3} \, &;& \, 
t^{22,23}=t^{23,22}=-\frac{1}{6} \, ; \nonumber \\
t^{23,33}=t^{33,23}=-\frac{1}{6} \, &;& \, 
t^{23,11} =t^{11,23}=\frac{1}{3}\,.
\ea
The matrix $\Delta_{ij}$ is defined as
\be
\Delta_{ij} \equiv u \lambda_{ij} u^\dagger\,,\quad
\left(\lambda_{ij}\right)_{ab} \equiv \delta_{ia} \, \delta_{jb}\,.
\ee
We only quote the $|\Delta S|=1$ part here but the others can be obtained
by changing the 
indices in $\Delta_{32}$ and $t^{ij,kl}$
appropriately. The parts with $G_8 $ and $G_8^\prime$
transforms as an octet under
$SU(3)_L$ while the $G_{27}$ part transforms as a 27 under the same group.
The term with $G_8^\prime$ is often referred to as the weak mass term. 

The coefficient $C$ is defined such that in the chiral
and large $N_c$ limits $G_8 = G_{27} =1$,
\be
C= -\frac{3}{5} \, \frac{G_F}{\sqrt 2} V_{ud} \, V_{us}^* \, .
\ee
The correspondence with the parameters $c_2$ and $c_3$ of \cite{KMW1,KMW2}
is
\be
c_2 = C F_0^4 G_8\,\quad c_3 = -\frac{1}{6} C  F_0^4 G_{27}\,.
\ee

The NLO nonleptonic weak Lagrangian was first worked out in \cite{KMW1}
but the basis given there was redundant. Subsequent work on a less redundant
basis was \cite{Esposito}. A fully nonredundant basis for the octet part
was presented in \cite{EKW}. We will use the basis of \cite{EKW}
for the octet part and use the notation of \cite{KMW1} for
the 27 part, but keep only the nonredundant terms as derived
in \cite{Esposito}. The NLO weak ChPT Lagrangian, quoting only the terms
relevant for $K\to2\pi$ and $K\to3\pi$ decays, is
\ba
\label{LW4}
{\cal L}^{(4)}_{\Delta S=1} &=& C \, F_0^2 \, \Bigg\{G_8\Big[
N_1 {\cal O}^8_1 + N_2 {\cal O}^8_2 + N_3 {\cal O}^8_3 + 
N_4 {\cal O}^8_4 
N_5 {\cal O}^8_5 + N_{6} {\cal O}^8_{6} 
+ N_{7} {\cal O}^8_{7}
 \nonumber \\&&
 + N_{8} {\cal O}^8_{8} + N_{9} {\cal O}^8_{9} 
+ N_{10} {\cal O}^8_{10} + N_{11} {\cal O}^8_{11} + 
N_{12} {\cal O}^8_{12} + N_{13} {\cal O}^8_{13} \Big]
\nonumber \\&&
+  G_{27} \,\Big[
D_1 {\cal O}^{27}_1 + D_2 {\cal O}^{27}_2 +
D_{4} {\cal O}^{27}_4 + D_{5} {\cal O}^{27}_5 
 + D_{6} {\cal O}^{27}_6 + D_{7} {\cal O}^{27}_7 + D_{26} {\cal O}^{27}_{26}
\nonumber \\&& 
+ D_{27} {\cal O}^{27}_{27} + D_{28} {\cal O}^{27}_{28}
+  D_{29} {\cal O}^{27}_{29}
+ D_{30} {\cal O}^{27}_{30} + D_{31} {\cal O}^{27}_{31} \Big]\Bigg\}
+ \mbox{h.c.}\,. 
\ea
The octet operators are
\ba
{\cal O}^8_1 
&=& \langle\Delta_{32} u_{\mu} u^{\mu} u_{\nu} u^{\nu} \rangle \,,
\nonumber \\
{\cal O}^8_2 
&=& \langle\Delta_{32} u_{\mu} u_{\nu} u^{\nu} u^{\mu} \rangle \,,
\nonumber \\ 
{\cal O}^8_3 
&=& \langle\Delta_{32} u_{\mu} u_{\nu}\rangle\langle u^{\mu} u^{\nu} 
\rangle \,,
\nonumber \\
{\cal O}^8_4 
&=& \langle\Delta_{32} u_{\mu}\rangle\langle u_{\nu} u^{\mu} u^{\nu} 
\rangle\,, 
\nonumber \\
{\cal O}^8_5 
&=& \langle\Delta_{32} (\chi_+ u_{\mu} u^{\mu}+u_{\mu} u^{\mu} \chi_+) 
\rangle \,,
\nonumber \\
{\cal O}^8_6 
&=& \langle\Delta_{32} u_{\mu}\rangle\langle u^{\mu} \chi_+ \rangle \,,
\nonumber \\
{\cal O}^8_7 
&=& \langle\Delta_{32} \chi_+\rangle\langle u_{\mu} u^{\mu} \rangle \,,
\nonumber \\
{\cal O}^8_8 
&=& \langle\Delta_{32} u_{\mu} u^{\mu}\rangle\langle\chi_+ \rangle \,,
\nonumber \\
{\cal O}^8_9 
&=& \langle\Delta_{32} (\chi_- u_{\mu} u^{\mu}-u_{\mu} u^{\mu} \chi_-)
 \rangle \,,
\nonumber \\
{\cal O}^8_{10} 
&=& \langle\Delta_{32} \chi_+ \chi_+ \rangle \,,
\nonumber \\
{\cal O}^8_{11} 
&=& \langle\Delta_{32} \chi_+\rangle\langle\chi_+ \rangle \,,
\nonumber \\
{\cal O}^8_{12} 
&=& \langle\Delta_{32} \chi_- \chi_- \rangle \,,
\nonumber \\
{\cal O}^8_{13} 
&=& \langle\Delta_{32} \chi_-\rangle\langle\chi_- \rangle \,.
\ea
The 27 operators are
\ba
{\cal O}^{27}_1 
&=& t^{ij,kl}  \langle \Delta_{ij} \chi_+\rangle
\langle\Delta_{kl} \chi_+\rangle \,,
\nonumber \\
{\cal O}^{27}_2 
&=& t^{ij,kl}  \langle\Delta_{ij} \chi_-\rangle
\langle\Delta_{kl} \chi_- \rangle \,,
\nonumber \\
{\cal O}^{27}_4 
&=& t^{ij,kl}  \langle\Delta_{ij} u_{\mu}\rangle
\langle\Delta_{kl} (u^{\mu} \chi_++\chi_+ u^{\mu}) \rangle \,,
\nonumber \\
{\cal O}^{27}_5 
&=& t^{ij,kl}  \langle\Delta_{ij} u_{\mu}\rangle
\langle\Delta_{kl} (u^{\mu} \chi_--\chi_- u^{\mu}) \rangle\,, 
\nonumber \\
{\cal O}^{27}_6 
&=& t^{ij,kl}  \langle\Delta_{ij} \chi_+\rangle
\langle\Delta_{kl} u_{\mu} u^{\mu} \rangle \,,
\nonumber \\
{\cal O}^{27}_7 
&=& t^{ij,kl}  \langle\Delta_{ij} u_{\mu}\rangle
\langle\Delta_{kl} u^{\mu}\rangle\langle\chi_+ \rangle \,,
\nonumber \\
{\cal O}^{27}_{26} 
&=& t^{ij,kl}  \langle\Delta_{ij} u_{\mu} u^{\mu}\rangle
\langle\Delta_{kl} u_{\nu} u^{\nu} \rangle\,,
\nonumber \\
{\cal O}^{27}_{27} 
&=& t^{ij,kl}  \langle\Delta_{ij} (u_{\mu} u_{\nu}+u_{\nu} u_{\mu})\rangle
\langle\Delta_{kl} 
( u^{\mu} u^{\nu} + u^{\nu} u^{\mu} ) \rangle\,,
\nonumber \\
{\cal O}^{27}_{28} 
&=& t^{ij,kl}  \langle\Delta_{ij} (u_{\mu} u_{\nu}-u_{\nu} u_{\mu})\rangle
\langle \Delta_{kl} 
( u^{\mu} u^{\nu} - u^{\nu} u^{\mu} )\rangle\,,
\nonumber \\
{\cal O}^{27}_{29} 
&=& t^{ij,kl}  \langle\Delta_{ij} u_{\mu}\rangle
\langle\Delta_{kl} u_{\nu} u^{\mu} u^{\nu} \rangle\,,
\nonumber \\
{\cal O}^{27}_{30} 
&=& t^{ij,kl}  \langle\Delta_{ij} u_{\mu}\rangle
\langle\Delta_{kl} (u^{\mu} u_{\nu} u^{\nu}+u_{\nu} u^{\nu} u^{\mu}) \rangle\,,
\nonumber \\
{\cal O}^{27}_{31} 
&=& t^{ij,kl}  \langle\Delta_{ij} u_{\mu}\rangle
\langle\Delta_{kl} u^{\mu}\rangle\langle u_{\nu} u^{\nu} \rangle\,.
\ea
Notice that some have the opposite sign compared to \cite{KMW1},
because of the difference in the definition of 
$\chi_i$ and $P$ of that reference.
The basis used in \cite{BPP} for the octet case is slightly different
but related to the one here by
\be
\begin{array}[b]{lll}
N_5= E_{10}-E_{11} \,; \quad&  N_6= E_{11} + 2 E_{12} \, ;\quad& 
N_7=\frac{1}{2} E_{11} + E_{13} \, ;\\
 N_8= E_{11} \, ; &
N_9=E_{15} \, ; &
 N_{10} = E_1-E_5 \, ;\\
N_{11}=E_2 \, ;&
 N_{12}= -E_3 + E_5 \, ; &
N_{13}=E_4 \, ; \\
 N_{36}=E_5 \, .
\end{array}
\ee
Notice that there is a small misprint in the relation between the
$E_i$ and $N_i$ in (2.19) of \cite{BPP} in the relation between $N_{13}$
and $E_4$.

The infinities appearing in the loop diagrams are canceled by
replacing the coefficients in (\ref{LS4}) and (\ref{LW4}) by
the renormalized coefficients and a subtraction part.
The infinities needed in the strong sector were calculated first
in \cite{GL2} and those for the weak sector
in \cite{KMW1} and confirmed in \cite{Esposito}.
For the terms in Eqs.~(\ref{LS4}) and (\ref{LW4}) they are all of the type
\be
\label{definf}
X_i = (e^c\mu)^{-2\epsilon}
\left(X_i^r + x_i\frac{1}{16\pi^2 \epsilon}\right)\,,
\ee
with the dimension of space-time $d=4-2\epsilon$ and
\be
c = -\frac{1}{2}\left(\ln(4\pi) +\Gamma^\prime(1)+1\right)\,,
\ee
for $X=L,N,D $. The coefficients are listed in Table \ref{tabinf}
\cite{GL2,KMW1,Esposito}.
\begin{table}
\begin{center}
\begin{tabular}{|cc|cc|cc|}
\hline
$L_i$   &   $\ell_i$  &   $N_i$   &  $n_i$   &   $D_i$    &  $d_i$  \\
\hline
1       & $-3/64  $   & 1  &  $-1  $                           &1 &$1/12$\\
2	& $-3/32  $   & 2  &  $1/4 $			       &2 &$0   $\\
3	& $0      $   & 3  &  $0   $			       &4 &$-3/2$\\
4	& $-1/16  $   & 4  &  $-1/2$			       &5 &$-1/2$\\
5	& $-3/16  $   & 5  &  $-3/4 -3/8  \,(G_8^\prime/G_8)$  &6 &$3/4 $\\
6	& $-11/288$   & 6  &  $1/8 $			       &7 &$-1/2$\\
7	& $0      $   & 7  &  $9/16 -1/4  \,(G_8^\prime/G_8)$  &26&$1/2 $\\
8	& $-5/96  $   & 8  &  $1/4  $			       &27&$1/4 $\\
9	& $-1/8   $   & 9  &  $-3/8 +3/8  \,(G_8^\prime/G_8)$  &28&$5/6 $\\
10      & $1/8    $   &	10 &  $-1/3 -5/24 \,(G_8^\prime/G_8)$  &29&$-19/$\\
        &      	      & 11 &  $13/36-11/36\,(G_8^\prime/G_8)$  &30&$-5/3$\\
        &      	      & 12 &  $5/24 -5/24 \,(G_8^\prime/G_8)$  &31&$0   $\\
        &             & 13 &  $0 $                             &  &      \\
\hline
\end{tabular}
\end{center}
\caption{The coefficients of the subtraction of the infinite
parts defined in Eq. (\ref{definf}).\label{tabinf}}
\end{table}

\section{Kinematics and Isospin}
\label{kinematics}

There are five CP-conserving decays of the type $K\to3\pi$:
\ba
\label{defdecays}
K_L(k)&\to&\pi^0(p_1)\pi^0(p_2)\pi^0(p_3)\,,\quad [A^L_{000}]\,,\nonumber\\
K_L(k)&\to&\pi^+(p_1)\pi^-(p_2)\pi^0(p_3)\,,\quad [A^L_{+-0}]\,,\nonumber\\
K_S(k)&\to&\pi^+(p_1)\pi^-(p_2)\pi^0(p_3)\,,\quad [A^S_{+-0}]\,,\nonumber\\
K^+(k)&\to&\pi^0(p_1)\pi^0(p_2)\pi^+(p_3)\,,\quad [A_{00+}]\,,\nonumber\\
K^+(k)&\to&\pi^+(p_1)\pi^+(p_2)\pi^-(p_3)\,,\quad [A_{++-}]\,,
\ea
where we have indicated the four momentum defined for each particle
and the symbol we will use for the amplitude.
The $K^+$ decays have an obvious counterpart in $K^-$ decays.

The kinematics is normally treated using
\be
s_1 = \left(k-p_1\right)^2\,,\quad
s_2 = \left(k-p_2\right)^2\,,\quad
s_3 = \left(k-p_3\right)^2\,.
\ee
The amplitudes are often expanded in terms of the Dalitz plot variables
$x$ and $y$ defined via
\be
\label{defsi}
\label{defxyexp}
x = \frac{s_2-s_1}{m_{\pi^+}^2}\,,
\quad
y = \frac{s_3-s_0}{m_{\pi^+}^2}\,,
\quad
s_0 = \frac{1}{3}\left(m_K^2+m_{\pi^1}^2+m_{\pi^2}^2+m_{\pi^3}^2\right)\,,
\ee
where the kaon mass and the pion masses in $s_0$ are those from the particles
appearing in the decay under consideration.

The amplitudes for
 $K_L\to\pi^+\pi^-\pi^0,K^+\to\pi^+\pi^+\pi^-$ and $K^+\to\pi^0\pi^0\pi^+$
are symmetric under the interchange of the first two pions because of
CP or Bose-symmetry. The amplitude for  $K_L\to\pi^0\pi^0\pi^0$ is
obviously symmetric
under the interchange of all three final state particles and the one for
$K_S\to\pi^+\pi^-\pi^0$ is antisymmetric under the interchange of $\pi^+$ and
$\pi^-$ because of CP.

Isospin invariance does give some constraints on the
amplitudes \cite{Zemach,Weinberg2,DAmbrosio}. These can be written as
\ba
\label{iso}
A^L_{000}(s_1,s_2,s_3) &=& 3 A_n(s_1,s_2,s_3) \,,\nonumber\\
A^L_{+-0}(s_1,s_2,s_3) &=& A_n(s_1,s_2,s_3)-B_n(s_1,s_2,s_3) \,,\nonumber\\
A^S_{+-0}(s_1,s_2,s_3) &=& C_0(s_1,s_2,s_3)
      +\frac{2}{3}\left[B_t(s_3,s_2,s_1)-B_t(s_3,s_1,s_2)\right]\,,\nonumber\\
A_{00+}(s_1,s_2,s_3) &=& A_c(s_1,s_2,s_3)-B_c(s_1,s_2,s_3)
+B_t(s_1,s_2,s_3) \,,\nonumber\\
A_{++-}(s_1,s_2,s_3) &=& 2A_c(s_1,s_2,s_3)+B_c(s_1,s_2,s_3)
+B_t(s_1,s_2,s_3) \,.
\ea
The functions $A_{c,n}$ are fully symmetric in $s_1,s_2,s_3$. $C_0$ is
fully antisymmetric
and $B_{c,n,t}(s_1,s_2,s_3)$ is symmetric under the interchange
$s_1\leftrightarrow s_2$ and satisfies the relation
\be
B_i(s_1,s_2,s_3)+B_i(s_2,s_3,s_1)+B_i(s_3,s_1,s_2) = 0\,.
\ee
$A_{c,n}$ and $B_{c,n}$ belong to the $I=1$ final state, $B_t$ to the $I=2$
final state
and $C_0$ to the $I=0$ final state.

A further simplification can be obtained by observing that the imaginary parts
of loop diagrams in ChPT do not contain $\ell\ge 2$ until at least 
${\cal O}(p^8)$.
This allows to rewrite the amplitudes in a way that only contains functions of
single variables and is fully correct up to  ${\cal O}(p^6)$. The underlying
arguments are the same as the ones used for a similar decomposition for
$\pi\pi$ scattering by \cite{knechtetal}.
\ba
\label{defMi}
A^L_{000}(s_1,s_2,s_3) &=& M_0(s_1)+M_0(s_2)+M_0(s_3) \,,
\nonumber\\
A^L_{+-0}(s_1,s_2,s_3) &=& M_1(s_3)+M_2(s_1)+M_2(s_2)+M_3(s_1)(s_2-s_3)
+M_3(s_2)(s_1-s_3)
\,,\nonumber\\
A^S_{+-0}(s_1,s_2,s_3) &=& M_4(s_1)-M_4(s_2)+M_5(s_1)(s_2-s_3)
-M_5(s_2)(s_1-s_3)
\nonumber\\&&
+M_6(s_3)(s_1-s_2)\,,
\nonumber\\
A_{00+}(s_1,s_2,s_3) &=&  M_7(s_3)+M_8(s_1)+M_8(s_2)+M_9(s_1)(s_2-s_3)
+M_9(s_2)(s_1-s_3)\,,
\nonumber\\
A_{++-}(s_1,s_2,s_3) &=&  M_{10}(s_3)+M_{11}(s_1)+M_{11}(s_2)
+M_{12}(s_1)(s_2-s_3)
+M_{12}(s_2)(s_1-s_3) \,.
\nonumber\\
\ea
The expressions for the various amplitudes become significantly shorter
when written
explicitly in this form.
Comparing (\ref{iso}) and (\ref{defMi}) gives
\ba
A_n(s_1,s_2,s_3)&=& \sum_{i=1,3}
\frac{1}{3} M_0(s_i)\,,\nonumber\\
A_c(s_1,s_2,s_3)&=& \sum_{i=1,3} \frac{1}{3}\left( M_7(s_i)
+ 2 M_8(s_i)\right)\,,\nonumber\\
B_n(s_1,s_2,s_3)&=&  B_{n1}(s_1)+B_{n1}(s_2)-2B_{n1}(s_3)+B_{n2}(s_1)(s_2-s_3)
+B_{n2}(s_2)(s_1-s_3)\,,
\nonumber\\
B_{n1}(s) &=& \frac{1}{3}\left( M_1(s)-M_2(s)\right)\,,
\nonumber\\
B_{n2}(s) &=&-M_3(s)\,,
\nonumber\\
B_c(s_1,s_2,s_3)&=&  B_{c1}(s_1)+B_{c1}(s_2)-2B_{c1}(s_3)+B_{c2}(s_1)(s_2-s_3)
+B_{c2}(s_2)(s_1-s_3)\,,
\nonumber\\
B_{c1}(s) &=& \frac{1}{6}\left(M_7(s)-M_8(s)-M_{10}(s)+M_{11}(s)\right)\,
\nonumber\\
B_{c2}(s) &=& \frac{1}{2}\left(-M_9(s)+M_{12}(s)\right)\,,
\nonumber\\
B_t(s_1,s_2,s_3)&=&  B_{t1}(s_1)+B_{t1}(s_2)-2B_{t1}(s_3)+B_{t2}(s_1)(s_2-s_3)
+B_{t2}(s_2)(s_1-s_3)\,,
\nonumber\\
B_{t1}(s) &=& \frac{1}{6}\left(-M_7(s)+M_8(s)-M_{10}(s)+M_{11}(s)\right)\,
\nonumber\\
B_{t2}(s) &=&\frac{1}{2}\left(M_9(s)+M_{12}(s)\right)\,,
\nonumber\\
C_0(s_1,s_2,s_3) &=& C_{01}(s_3)(s_1-s_2)+ C_{01}(s_1)(s_2-s_3)+
C_{01}(s_2)(s_3-s_1)\,,
\nonumber\\
C_{01}(s) &=& \frac{1}{3}\left(2 M_5(s)+M_6(s)\right)
\ea
and the relations
\ba
\label{isorelations}
M_0(s) &=& M_1(s)+2M_2(s)\,,
\nonumber\\
2M_7(s)+4M_8(s) &=& M_{10}(s)+2M_{11}(s)\,,
\nonumber\\
M_4(s) &=& \frac{1}{3}\left(M_7(s)-M_8(s)+M_{10}(s)-M_{11}(s)\right)\,
\nonumber\\
M_5(s)-M_6(s)&=& M_9(s)+M_{12}(s)\,.
\ea
The split-up between
the polynomial parts in (\ref{defMi}) has some ambiguity as discussed
in App.~\ref{App:ambiguity}.
The same ambiguity appears
in the rewriting of the $B_{cnt}$ functions in single variable parts and can
affect the last two relations in (\ref{isorelations}).
The results given explicitly in App.~\ref{App:results}
are brought in a form satisfying the relations (\ref{isorelations})
using this freedom.

\section{Analytical Results}
\label{analytical}

\begin{figure}
\begin{center}
\setlength{\unitlength}{1pt}
\begin{picture}(260,100)
\SetScale{1.0}
\SetWidth{2}
\Line(0,50)(50,50)
\Line(50,50)(100,50)
\Line(50,50)(50,100)
\Line(50,50)(50,0)
\GBoxc(50,50)(5,5){0}
\Line(150,50)(180,50)
\Line(180,50)(210,50)
\GBoxc(180,50)(5,5){0}
\Line(210,50)(210,100)
\Line(210,50)(210,0)
\Line(210,50)(260,50)
\Vertex(210,50){5}
\end{picture}
\end{center}
\caption{The tree level diagrams for $K\to3\pi$. A filled square is a vertex
from ${\cal L}_{W2}$ and a filled circle from ${\cal L}_{S2}$.
\label{figtree}}
\end{figure}
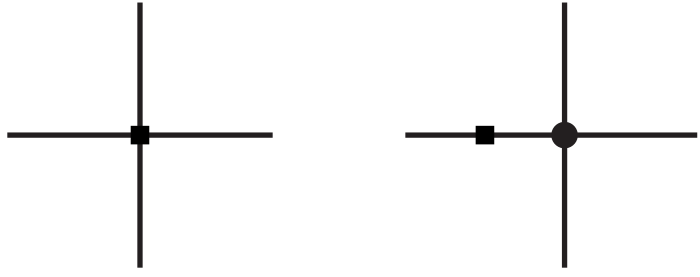

The lowest order result is well known and follows from the diagrams in
Fig.\ref{figtree}.
The Feynman amplitude at tree level for $A^L_{000}$ is
\be
\label{M0tree}
\left.M_0(s)\right|_{p^2} = i\frac{C F_0^4}{F_\pi^3 F_K}\,
\frac{1}{3}\left(G_8-G_{27}\right) m_K^2\,.
\ee
A choice for $A^L_{+-0}$ compatible with the relations
 (\ref{isorelations}) and (\ref{M0tree}) is
\ba
\left.M_1(s)\right|_{p^2}  &=&i\frac{C F_0^4}{F_\pi^3 F_K}\,
\frac{1}{3}\left(G_8-G_{27}\right) m_K^2\,,
\nonumber\\
\left.M_2(s)\right|_{p^2} &=& 0\,,
\nonumber\\
\left.M_3(s)\right|_{p^2} &=& i\frac{C F_0^4}{F_\pi^3 F_K}\,
\left\{
-\frac{G_8}{3}
+\frac{G_{27}}{6}\,\frac{-3m_K^2+8m_\pi^2}{m_K^2-m_\pi^2}
\right\}
\ea
The tree level result for $A^S_{+-0}$ allows very much of reshuffling
between $M_4$, $M_5$ and $M_6$. A choice is
\ba
\left.M_4(s)\right|_{p^2} &=& i\frac{C F_0^4}{F_\pi^3 F_K}\,
\frac{G_{27}}{m_K^2-m_\pi^2}\,\frac{5}{6}\left(2 m_\pi^2-3 m_K^2\right)s\,,
\nonumber\\
\left.M_5(s)\right|_{p^2} &=&  i\frac{C F_0^4}{F_\pi^3 F_K}\,
\frac{G_{27}}{m_K^2-m_\pi^2}\,\frac{5}{6}\left(2 m_\pi^2-3 m_K^2\right)\,,
\nonumber\\
\left.M_6(s)\right|_{p^2} &=&i\frac{C F_0^4}{F_\pi^3 F_K}\,
 \frac{G_{27}}{m_K^2-m_\pi^2}\frac{5}{6}\left(2m_\pi^2-3m_K^2\right)\,.
\ea
$A_{00+}$ leads to
\ba
\left.M_7(s)\right|_{p^2} &=&i\frac{C F_0^4}{F_\pi^3 F_K}\,
\Bigg\{\frac{G_8}{2}\left(m_\pi^2+m_K^2-s\right)
\nonumber\\&&
+\frac{G_{27}}{m_K^2-m_\pi^2}
\left(\frac{-5}{3}m_\pi^2m_K^2+\frac{13}{6}m_\pi^4-\frac{1}{2}m_K^4
+\frac{7}{6}m_\pi^2 s-\frac{17}{6}m_K^2 s\right)\Bigg\}\,,
\nonumber\\
\left.M_8(s)\right|_{p^2} &=& -\left.M_7(s)\right|_{p^2} +
i\frac{C F_0^4}{F_\pi^3 F_K}\,
 \frac{G_{27}}{m_K^2-m_\pi^2}\frac{5}{6}\left(2m_\pi^2-3m_K^2\right)s\,,
\nonumber\\
\left.M_9(s)\right|_{p^2} &=& 0\,.
\ea
The amplitude for $A_{++-}$ can be written as
\ba
\left.M_{10}(s)\right|_{p^2} &=&i\frac{C F_0^4}{F_\pi^3 F_K}\,
\left\{G_8\left(s-m_\pi^2-m_K^2\right)
+G_{27}\left(-\frac{13}{3}\,s+m_K^2+\frac{13}{3}m_\pi^2 \right)\right\}\,,
\nonumber\\
\left.M_{11}(s)\right|_{p^2} &=& 0\,,
\nonumber\\
\left.M_{12}(s)\right|_{p^2} &=& 0\,.
\ea

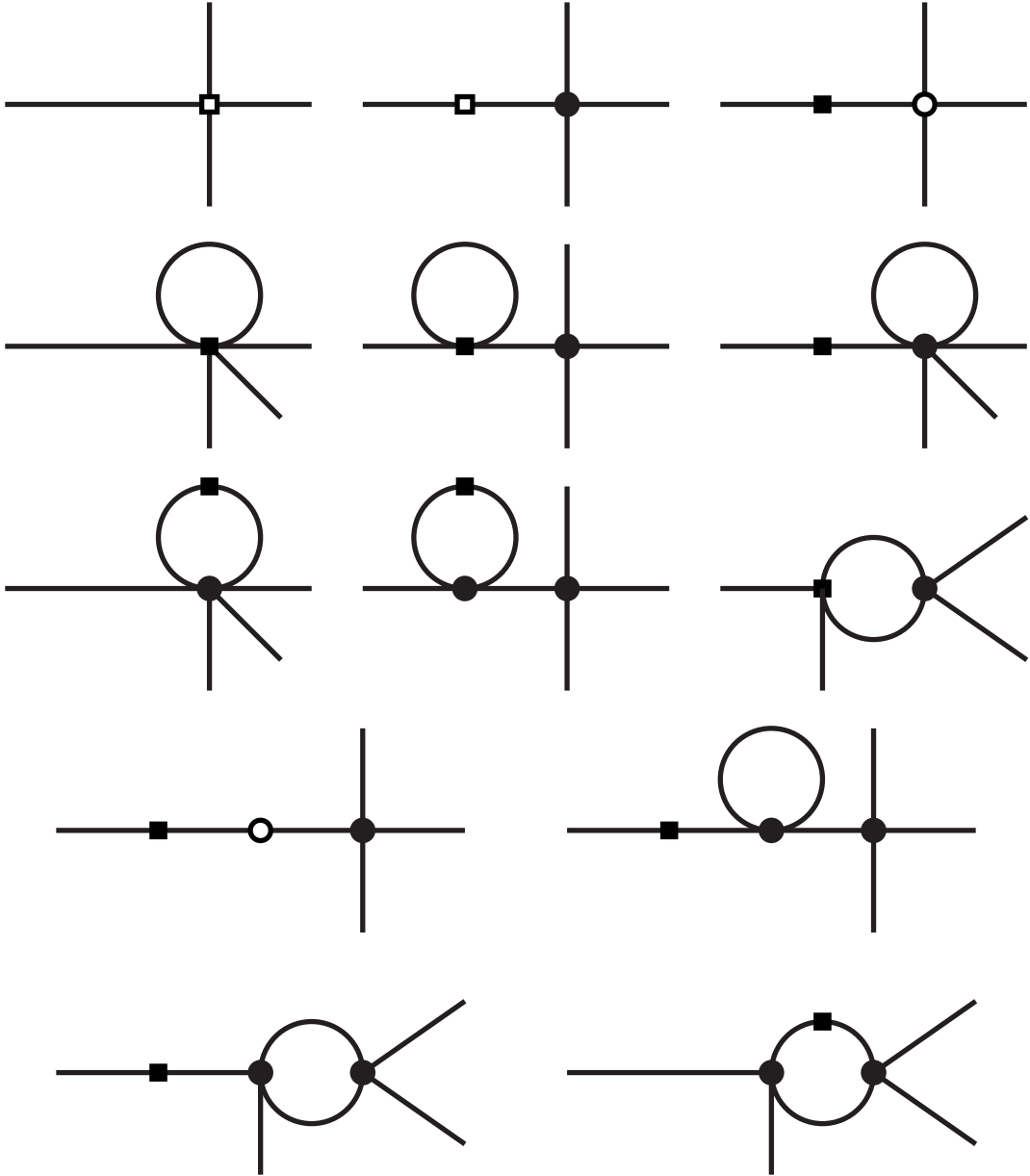
\begin{figure}
\begin{center}
\setlength{\unitlength}{1pt}
\begin{picture}(400,80)
% Diagram 3
\SetScale{1.0}
\SetWidth{2}
\Line(80,40)(120,40)
\Line(80,40)(80,80)
\Line(80,40)(80,0)
\Line(0,40)(80,40)
\GBoxc(80,40)(6,6){1}
% Diagram 4
\Vertex(220,40){5}
\Line(220,40)(260,40)
\Line(220,40)(220,80)
\Line(220,40)(220,0)
\Line(140,40)(220,40)
\GBoxc(180,40)(6,6){1}
% Diagram 5
\Line(360,40)(400,40)
\Line(360,40)(360,80)
\Line(360,40)(360,0)
\Line(280,40)(360,40)
\GBoxc(320,40)(5,5){0}
\GCirc(360,40){4}{1}
\end{picture}
\\[0.5cm]
\setlength{\unitlength}{1pt}
\begin{picture}(400,80)
\SetScale{1.0}
\SetWidth{2}
% Diagram 7
\Line(80,40)(120,40)
\Line(80,40)(80,80)
\Line(80,40)(80,0)
\Line(0,40)(80,40)
\BCirc(80,60){20}
\Line(80,40)(108,12)
\GBoxc(80,40)(5,5){0}
% Diagram 8
\Vertex(220,40){5}
\Line(220,40)(260,40)
\Line(220,40)(220,80)
\Line(220,40)(220,0)
\Line(140,40)(220,40)
\BCirc(180,60){20}
\GBoxc(180,40)(5,5){0}
% Diagram 9
\Line(360,40)(400,40)
\Line(360,40)(360,80)
\Line(360,40)(360,0)
\Line(280,40)(360,40)
\GBoxc(320,40)(5,5){0}
\BCirc(360,60){20}
\Line(360,40)(388,12)
\Vertex(360,40){5}
\end{picture}
\\[0.5cm]
\setlength{\unitlength}{1pt}
\begin{picture}(400,80)
\SetScale{1.0}
\SetWidth{2}
% Diagram 11
\Line(80,40)(120,40)
\Line(80,40)(80,80)
\Line(80,40)(80,0)
\Line(0,40)(80,40)
\BCirc(80,60){20}
\Line(80,40)(108,12)
\Vertex(80,40){5}
\GBoxc(80,80)(5,5){0}
% Diagram 12
\Vertex(220,40){5}
\Line(220,40)(260,40)
\Line(220,40)(220,80)
\Line(220,40)(220,0)
\Line(140,40)(220,40)
\BCirc(180,60){20}
\Vertex(180,40){5}
\GBoxc(180,80)(5,5){0}
% Diagram 13
\BCirc(340,40){20}
\Line(280,40)(320,40)
\GBoxc(320,40)(5,5){0}
\Line(320,40)(320,0)
\Vertex(360,40){5}
\Line(360,40)(400,68)
\Line(360,40)(400,12)
\end{picture}
\\[0.5cm]
\setlength{\unitlength}{1pt}
\begin{picture}(400,80)
\SetScale{1.0}
\SetWidth{2}
% Diagram 6
\Line(20,40)(60,40)
\Line(60,40)(100,40)
\Line(100,40)(140,40)
\Line(140,40)(180,40)
\Line(140,40)(140,80)
\Line(140,40)(140,0)
\Vertex(140,40){5}
\GCirc(100,40){4}{1}
\GBoxc(60,40)(5,5){0}
% Diagram 10
\Line(220,40)(260,40)
\Line(260,40)(300,40)
\Line(300,40)(340,40)
\Line(340,40)(380,40)
\Line(340,40)(340,80)
\Line(340,40)(340,0)
\BCirc(300,60){20}
\Vertex(340,40){5}
\Vertex(300,40){5}
\GBoxc(260,40)(5,5){0}
\end{picture}
\\[0.5cm]
\setlength{\unitlength}{1pt}
\begin{picture}(400,80)
\SetScale{1.0}
\SetWidth{2}
% Diagram 14
\Line(20,40)(60,40)
\Line(60,40)(100,40)
\Line(140,40)(180,68)
\Line(140,40)(180,12)
\Line(100,40)(100,0)
\BCirc(120,40){20}
\Vertex(140,40){5}
\Vertex(100,40){5}
\GBoxc(60,40)(5,5){0}
% Diagram 15
\Line(300,40)(300,0)
\Line(220,40)(300,40)
\BCirc(320,40){20}
\Vertex(300,40){5}
\Vertex(340,40){5}
\GBoxc(320,60)(5,5){0}
\Line(340,40)(380,68)
\Line(340,40)(380,12)
\end{picture}
\end{center}
\caption{The diagrams of order $p^4$.
An open square is a vertex
from ${\cal L}_{W4}$, an open circle a vertex from ${\cal L}_{S4}$,
a filled square a vertex
from ${\cal L}_{W2}$ and a filled circle a vertex from ${\cal L}_{S2}$.
Diagrams contributing to strong interaction wave function
renormalization are not shown. 
\label{figp4}}
\end{figure}

The result at order $p^4$ is significantly longer. The diagrams that contribute
are shown in Fig. \ref{figp4}. Using the decomposition of all the
amplitudes in the $M_i(s)$, and using the freedom in choosing the functions
$M_i(s)$, relatively compact expressions can be obtained and we
have given them explicitly in App.~\ref{App:results}.
Our results for $K\to\pi\pi$ are in full agreement with \cite{BPP}.

In order to do the data fits it is important to see how many combinations
of the 25 unknown  parameters appearing in (\ref{LW4}) actually matter.
Taking the various independent coefficients from the amplitudes $M_i(s)$,
it turns out that only 11 linear combinations of the 25 parameters
show up. Of these seven
appear in the amplitudes multiplied by coefficients of order $m_K^4$
while the remaining four have coefficients of order $m_\pi^2 m_K^2$.

The measurable combinations are displayed in table \ref{tab:Ki}. Any
linear combination of them is of course also allowed. The choice
displayed in the table was made to have all combinations to be orthogonal
to the non-measurable ones when possible. The remainder
was then chosen simply to have somewhat compact expressions for their
contributions to the $M_i(s)$ and the $A_0$ and $A_2$ amplitude
for $K\to\pi\pi$.
$\tilde K_4$ is the combination
multiplying $m_K^4$ in the  $A_2$ amplitude. $\tilde K_1$
is the octet part of the $A_0$ amplitude proportional to $m_K^4$
plus the part of the 27 multiplying $m_K^4$
that cannot be written in terms of $\tilde K_4$.
$\tilde K_{10}$ and $\tilde K_{8}$ are the equivalent parts of the
coefficients of $m_\pi^4$ in $A_2$ and $A_0$.
The numbering is chosen such that $\tilde K_1$ to $\tilde K_7$ 
contribute leading in $m_K^2$ and inside those categories the ones
with octet contributions are listed first.
There are 5 combinations with octet contributions of which three leading
in $m_K^2$.

\begin{table}
\begin{center}
\begin{tabular}{|c|c|}
\hline
$\tilde K_{1}$  &  $G_8\left( N_5^r-2 N_7^r+2 N_8^r+N_9^r \right)
         +G_{27}\left(-\frac{1}{2} D_6^r \right)$\\     
$\tilde K_{2}$  &  $G_8\left(N_1^r+N_2^r  \right)
     +G_{27}\left(\frac{1}{3} D_{26}^r -\frac{4}{3}D_{28}^r\right)$\\     
$\tilde K_{3}$  &  $G_8\left(N_3^r  \right)
        +G_{27}\left(\frac{2}{3}D_{27}^r+\frac{2}{3}D_{28}^r \right)$\\     
$\tilde K_{4}$  &  $G_{27}\left(D_{4}^r-D_{5}^r+4 D_{7}^r \right)$\\     
$\tilde K_{5}$  &  $G_{27}\left(D_{30}^r+D_{31}^r+2 D_{28}^r \right)$\\
$\tilde K_{6}$  &  $G_{27}\left(8 D_{28}^r-D_{29}^r+D_{30}^r \right)$\\
$\tilde K_{7}$  &  $G_{27}\left(-4 D_{28}^r+D_{29}^r \right)$\\
$\tilde K_{8}$  &  $G_8\left(2 N_5^r+4 N_7^r+N_8^r
        -2 N_{10}^r-4 N_{11}^r-2 N_{12}^r \right)
       +G_{27}\left(-\frac{2}{3} D_1^r+\frac{2}{3} D_6^r \right)$\\     
$\tilde K_{9}$  &  $G_8\left( N_5^r+N_8^r+N_9^r \right)
                   +G_{27}\left(-\frac{1}{6} D_6^r \right)$\\     
$\tilde K_{10}$ &  $G_{27}\left(2 D_2^r-2 D_4^r-D_7^r \right)$\\     
$\tilde K_{11}$ &  $G_{27}\left(D_7^r \right)$\\     
\hline
\end{tabular}
\end{center}
\caption{The independent linear combinations of the $N_i^r$ and $D_i^r$
that appear in the amplitudes.\label{tab:Ki}}
\end{table}

\begin{figure}
\begin{center}\includegraphics[width=0.8\textwidth]{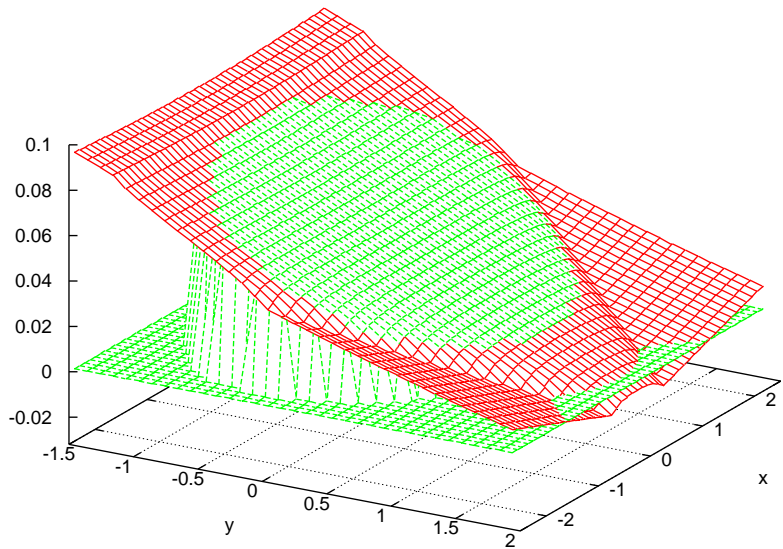}
\end{center}
\caption{\label{figphase} $\mathrm{Im}(A_{++-})$ as a function of 
$x$ and $y$ normalized to $-|A_{++-}(s_0,s_0,s_0)|$. The curve
with the plateau at zero indicates the phasespace. The thresholds
in the various $\pi\pi$ channels are clearly visible.}
\end{figure}

The amplitudes we calculated are at one loop in ChPT. They thus include
final state rescattering effects. A partial analysis of this type
of effects was done in \cite{DAmbrosio}. They derived the relations
\ba
\label{FSIrelations}
\left(\begin{array}{c}A_c(s_1,s_2,s_3)\\B_c(s_1,s_2,s_3)\end{array}\right)_{R}
&=& R(s_1,s_2,s_3)
\left(\begin{array}{c}A_c(s_1,s_2,s_3)\\B_c(s_1,s_2,s_3)\end{array}\right)_{NR}
\nonumber\\
\left(\begin{array}{c}A_n(s_1,s_2,s_3)\\B_n(s_1,s_2,s_3)\end{array}\right)_{R}
&=& R(s_1,s_2,s_3)
\left(\begin{array}{c}A_n(s_1,s_2,s_3)\\B_n(s_1,s_2,s_3)\end{array}\right)_{NR}
\nonumber\\
B_2(s_1,s_2,s_3)_R &=& \delta(s_1,s_2,s_3) B_2(s_1,s_2,s_3)_{NR}
\ea
for the amplitudes defined in (\ref{iso}). The subscript $R$ on the l.h.s.
means that rescattering effects have been included and $NR$ on the r.h.s.
that they are not included. The expressions for the 
matrix function $R(s_1,s_2,s_3)$ and the function $\delta(s_1,s_2,s_3)$
to lowest order in ChPT can be found in \cite{DAmbrosio}. We have checked
numerically that our amplitudes satisfy those relations to the order required.
The approach to the various thresholds ($s_i\to4 m_\pi^2$) is as expected
to this order in ChPT. The effects of the thresholds can be seen
in Fig. \ref{figphase}. The plot shows $\mathrm{Im}(A_{++-})$ as a function of 
$x$ and $y$ normalized to $-|A_{++-}(s_0,s_0,s_0)|$

\section{Experimental Data and Fits}
\label{dataplusfits}

At the end of the seventies a review was written incorporating the then
finished precision experiments on the charged kaon decays \cite{Devlin}.
Since then there have been many new results on $K\to3\pi$ decays.
A recent attempt to refit the data using some of the partial numerics
of \cite{KMW2} is \cite{Cheshkov}. We have in this paper recalculated
the full expressions and will do a full fit to the data contrary to
\cite{Cheshkov}. 
We analyze the data in terms of the widths given in Table \ref{tab:widths}.
The numbers are taken from the review of particle
properties \cite{PDG2000}. We do not directly use the data for
$\Gamma^S_{+-0}$ but instead fit directly to the measured quantity $\lambda$
as discussed below.
\begin{table}
\begin{center}
\begin{tabular}{|c|c|c|}
\hline
Decay   & Width [GeV] & CHPT fit [GeV]\\
\hline
$K^+\to\pi^+\pi^0$     &$(1.1245\pm0.0078)\cdot10^{-17}$&$1.124\cdot10^{-17}$\\
$K_S\to\pi^0\pi^0$     &$(2.3124\pm0.0207)\cdot10^{-15}$&$2.312\cdot10^{-15}$\\
$K_S\to\pi^+\pi^-$     &$(5.0543\pm0.0211)\cdot10^{-15}$&$5.054\cdot10^{-15}$\\
\hline
$K_L\to\pi^0\pi^0\pi^0$&$(2.6901\pm0.0402)\cdot10^{-18}$&$2.655\cdot10^{-18}$\\
$K_L\to\pi^+\pi^-\pi^0$&$(1.5978\pm0.0283)\cdot10^{-18}$&$1.626\cdot10^{-18}$\\
$K_S\to\pi^+\pi^-\pi^0$&$(2.36\pm0.81)\cdot10^{-21}    $&$3.1\cdot10^{-21}  $\\
$K^+\to\pi^0\pi^0\pi^+$&$(0.9194\pm0.0426)\cdot10^{-18}$&$0.911\cdot10^{-18}$\\
$K^+\to\pi^+\pi^+\pi^-$&$(2.9706\pm0.0272)\cdot10^{-18}$&$2.973\cdot10^{-18}$\\
\hline  
\end{tabular}
\end{center}
\caption{The various decay widths from the PDG tables\cite{PDG2000}.}
\label{tab:widths}
\end{table}

In addition several slope parameters are measured in the various
decays. The distributions in the Dalitz plot are conventionally
described in terms of $x$ and $y$ defined in (\ref{defsi}).

The decay amplitudes squared are now expanded as
\be
\left|\frac{A(s_1,s_2,s_3)}{A(s_0,s_0,s_0)}\right|^2
= 1 + g y + h y^2 + k x^2
\ee
where we have used $CP$ invariance and the symmetries in the decays.
For $K_L\to\pi^0\pi^0\pi^0$ $g=0$ and $k=h/3$.
For $K_S\to\pi^+\pi^-\pi^0$ one defines the ratio $\lambda$ or measures
directly the expansion of the amplitude \cite{CPLEAR}
\ba
\lambda&=& \frac{\dsp\int_{y_{min}}^{y_{max}}dy\int_0^{x_{lim}(y)}dx
A^{L*}_{+-0} A^S_{+-0}}{\dsp\int_{y_{min}}^{y_{max}}dy\int_0^{x_{lim}(y)}dx
A^{L*}_{+-0} A^L_{+-0}}
\nonumber\\
A^S_{+-0} &=& \gamma_S x - \xi_S xy\,.
\ea
The various measurements are given in Table \ref{tab:dalitz}.
Experiments after 1980 have been cited explicitly. 
Earlier references can be found in the PDG\cite{PDG2000} or
in the comprehensive review  \cite{Devlin}. Notice that there are
significant discrepancies between the various experiments indicated
by the fairly large scale factors.
\begin{table}[t]
\begin{center}
\begin{tabular}{|c|c|c|c|c|c|}
\hline
Decay & Quantity &     & S & Ref. &CHPT fit\\
\hline
$K_L\to\pi^0\pi^0\pi^0$ & $h$ & $-0.0033\pm0.0013$  & & \cite{E731} & \\
                        & $h$ & $-0.0061\pm0.0010$  & & \cite{NA48} &\\
                        & $h$ & $-0.00506\pm0.00135$& 1.7 & average&$-$0.0072\\
\hline
$K_L\to\pi^+\pi^-\pi^0$ & $g$ & $0.678\pm0.008$ & 1.5 & \cite{CPLEAR,PDG2000}
  &0.677\\
    & $h$ & $0.076\pm0.006$ &       & \cite{CPLEAR,PDG2000}&0.085\\
    & $k$ & $0.0099\pm0.0015$ &     & \cite{CPLEAR,PDG2000}& 0.0055\\
\hline
$K_S\to\pi^+\pi^-\pi^0$ & Re($\lambda$) &$0.0316\pm0.0062$&&\cite{CPLEAR,Zou}
 &0.0359\\
                        & Im($\lambda$) &$-0.0088\pm0.0068$&&\cite{CPLEAR,Zou}
 &$-$0.003\\
                 & $\gamma_S$& $(3.3\pm0.5)\cdot10^{-8}$&& \cite{CPLEAR}
 &$3.4\cdot10^{-8}$\\
                 & $\xi_S$   & $(0.4\pm0.8)\cdot10^{-8}$&&\cite{CPLEAR}
 &$-0.2\cdot10^{-8}$\\
\hline
$K^{\pm}\to\pi^0\pi^0\pi^{\pm}$ & $g$ & $0.652\pm0.031$ & 2.7&
                \cite{Batusov,Bolotov,PDG2000}&0.638 \\
                        & $h$ & $0.057\pm0.018$ & 1.4 & 
                 \cite{Batusov,Bolotov,PDG2000}&0.074\\
                        & $k$ & $0.0197\pm0.0054$ &  & 
                 \cite{Batusov,PDG2000}&0.0045 \\
\hline
$K^+\to\pi^+\pi^+\pi^-$ & $g$ & $-0.2154\pm0.0035$ & 1.4 & \cite{PDG2000}
   &  $-$0.216\\
   & $h$ & $0.012\pm0.008$ & 1.4 & \cite{PDG2000} &0.012\\
   & $k$ & $-0.0101\pm0.0034$ & 2.1 & \cite{PDG2000}& $-$0.0052\\
\hline
$K^-\to\pi^-\pi^-\pi^+$ & $g$ & $-0.217\pm0.007$ & 2.5 & \cite{PDG2000} & \\
                        & $h$ & $0.010\pm0.006$ &    & \cite{PDG2000}  & \\
                        & $k$ & $-0.0084\pm0.0019$ & 2.1 & \cite{PDG2000} &\\
\hline
\end{tabular}
\end{center}
\caption{The measurements of the Dalitz plot distributions. Notice that
in many cases the scale factors calculated
with the PDG\cite{PDG2000} method are sizable. The references
refer to the most recent measurements and the PDG. S is the scale
factor calculated using the prescription of \cite{PDG2000}.}
\label{tab:dalitz}
\end{table}

The data have been analyzed previously in terms of expansions in $x$ and $y$
of the amplitudes.
\ba
\label{defab}
A^L_{000}&=& 3(\alpha_1+\alpha_3)
         +3 (\zeta_1-2\zeta_3)\left(y^2+\frac{1}{3}x^2\right)\,,
\nonumber\\
A^L_{+-0}&=& (\alpha_1+\alpha_3)-(\beta_1+\beta_3)y
         + (\zeta_1-2\zeta_3)\left(y^2+\frac{1}{3}x^2\right)
         +(\xi_1-2\xi_3)\left(y^2-\frac{1}{3}x^2\right) \,,
\nonumber\\
A^S_{+-0}&=& \frac{2}{3}\sqrt{3}\,\gamma_3 x- \frac{4}{3}\xi_3^\prime xy\,,
\nonumber\\
A_{00+}&=& \left(-\alpha_1+\frac{1}{2}\alpha_3\right)
           +\left(\beta_1-\frac{1}{2}\beta_3-\sqrt{3}\gamma_3\right)y
         + (-\zeta_1-\zeta_3)\left(y^2+\frac{1}{3}x^2\right)
\nonumber\\&&
         +(-\xi_1-\xi_3-\xi_3^\prime)\left(y^2-\frac{1}{3}x^2\right) \,,
\nonumber\\
A_{++-}&=& \left(-2\alpha_1+\alpha_3\right)
           +\left(-\beta_1+\frac{1}{2}\beta_3-\sqrt{3}\gamma_3\right)y
         + (-2\zeta_1-2\zeta_3)\left(y^2+\frac{1}{3}x^2\right)
\nonumber\\&&
         +(\xi_1+\xi_3-\xi_3^\prime)\left(y^2-\frac{1}{3}x^2\right) \,.
\ea

A main problem in dealing with the data fitting is the question of how
to deal with the isospin breaking effects. The phase-space itself
is very sensitive to the precise masses of the pions and kaons
which are used. There are several ways to deal with this problem.
One is to first fit the expressions (\ref{defab}) to the data
where $x$ and $y$ are defined with $s_0$ calculated with the physical
masses but otherwise the charged pion mass as in (\ref{defxyexp}).
The result of this fit is shown in Table \ref{tab:ab}. Here we have assumed
that the only complex phase appearing is the relative phase between $A_0$
and $A_2$. The errors quoted are the minuit errors. For the $K_S$-measurements
only $\mbox{Re}(\lambda)$ is included. Leaving it out changes the $\chi^2/DOF$
to 3.9/4 but all parameters stay within the errors given.
\begin{table}
\begin{center}
\begin{tabular}{|c|c|c|c|c|c|}
\hline
quantity & Ref. \cite{Devlin} & Ref. \cite{KMW2} & Our fit  & $p^2$ &
$\tilde K_i=0$\\
\hline
$|A_0|$             &$0.4687\pm0.0006$&$0.4699\pm0.0012$&$0.4622\pm0.0014$
&input&input\\
$|A_2|$             &$0.0210\pm0.0001$&$0.0211\pm0.0001$&$0.0212\pm0.0001$
&input&input\\
$\delta_2-\delta_0$ &$(-45.6\pm5)^{\rm o}$&$(-61.5\pm4)^{\rm o}$
&$(-58.2\pm4)^{\rm o}$& ---& ---\\
$\alpha_1$          &$91.4\pm0.24    $&$91.71\pm0.32   $&$93.16\pm0.36   $
&74.0(73.5)&59.4\\
$\alpha_3$          &$-7.14\pm0.36   $&$-7.36\pm0.47   $&$-6.72\pm0.46   $
&$-4.8(4.8)$&$-6.5$\\
$\beta_1$           &$-25.83\pm0.41  $&$-25.68\pm0.27  $&$-27.06\pm0.43  $
&$-17.7(16.2)$&$-21.9$\\
$\beta_3$           &$-2.48\pm0.48   $&$-2.43\pm0.41   $&$-2.22\pm0.47   $
&$-1.2(1.1)$&$-1.0$\\
$g_3$               &$2.51\pm0.36    $&$2.26\pm0.23    $&$2.95\pm0.32    $
&$2.3(2.1)$&$2.5$\\
$\zeta_1$           &$-0.37\pm0.11   $&$-0.47\pm0.15   $&$-0.40\pm0.19   $
&---&0.26\\
$\zeta_3$           & ---             &$-0.21\pm0.08   $&$-0.09\pm0.10   $
&---&$-0.01$\\
$\xi_1$             &$-1.25\pm0.12   $&$-1.51\pm0.30   $&$-1.83\pm0.30   $
&---&$-0.46$\\
$\xi_3$             & ---             &$-0.12\pm0.17   $&$-0.17\pm0.16   $
&---&$-0.01$\\
$\xi_3^\prime$      & ---             &$-0.21\pm0.51   $&$-0.56\pm0.42   $
&---&$-0.06$\\
\hline
$\chi^2/DOF$        &$ 12.8/3          $&$10.3/2       $&$5.4/5 $
&---&---\\
\hline
\end{tabular}
\end{center}
\caption{The values of the parameters in the expansion in $x$ and $y$
of the $K\to3\pi$ amplitudes and the $K\to\pi\pi$ amplitudes.
The $\chi^2$ quoted are for the fits with the then used data.
$|A_0|$ and $|A_2|$ are in units of $10^{-6}~GeV$. 
$\alpha_1$,\ldots,$\xi_3^\prime$  are in units of $10^{-8}$.
\label{tab:ab}}
\end{table}

In the column labeled $p^2$ in
Table \ref{tab:ab} we fixed $G_8$ and $G_{27}$ from the result for
$A_0$ and $A_2$ at tree level
and show the predictions for the linear slopes in the amplitude at this
order. The source of the differences with \cite{KMW2} is not obvious
but could be due to a different pion and kaon mass. The numbers in the table
were calculated using the charged pion and kaon masses, using the neutral
ones instead leads to the numbers in brackets. As can be seen a general
reasonable agreement within about 40\% is obtained.
The input here determines $G_8 = 10.4$ and $G_{27}=0.61$
when we used for $F_0$ the pion decay constant $F_\pi$
and used $F_\pi$ and $F_K$ in the tree level amplitude with the
physical values.

Going to the next order, we first show also in Table
\ref{tab:ab} in the column labeled $\tilde K_i = 0$ the $p^4$
results with $G_8$ and $G_{27}$ fixed to fit $|A_0|$ and $|A_2|$.
These were calculated with charged pion and kaon mass.
The input here determines $G_8 = 5.47$ and $G_{27}=0.39$.
The change compared to the previous is an indication of the size of the
$p^4$ effect. It should be kept in mind that we have chosen
to normalize the lowest order including one factor of $1/F_K$, changing that
to $1/F_\pi$ changes the relative size of the $p^2$ and $p^4$ contributions
significantly. There is as well a mild dependence on the value of the $L_i^r$
used. We use in general a scale of $\mu=0.77$~GeV and the $L_i^r$ values
of the one-loop fit of \cite{ABT2}, the $p^4$ fit 
including the latest $K_{\ell4}$ data. These values are
(all at $\mu=0.77$~GeV.)
\ba
&\dsp  L_1^r =     0.38\cdot10^{-3},\quad 
  L_2^r =     1.59\cdot10^{-3},\quad
  L_3^r =    -2.91\cdot10^{-3} ,\quad
  L_4^r =      0, &
\nonumber\\
&\dsp  L_5^r =     1.46\cdot10^{-3},\quad 
  L_6^r =      0    ,\quad
  L_7^r =    -0.49\cdot10^{-3},\quad  
  L_8^r =     1.0\cdot10^{-3} \,.
&
\ea

It is not possible to determine all 13 quantities
$G_8$, $G_{27}$ and $\tilde K_1$,\ldots,$\tilde K_{11}$. In principle
it cannot be done if we try to fit to the 12 quantities
listed in Table \ref{tab:ab},\footnote{$\delta_2-\delta_0$ cannot be
fitted at this order.} but in practice the situation is even worse,
simply fitting the possible parameters without constraints leads to fits
with negligible tree level contributions. We have therefore chosen a two-step
process, first we study the variation of fitting subsets of
$G_8$, $G_{27}$ and $\tilde K_1$,\ldots,$\tilde K_{11}$
to $|A_0|$,\ldots,$\xi_3^\prime$. This allows for a more direct study
of the various dependencies on parameters.
Finally we present the direct fit
where we compare directly to the data, but allowing in addition for
a variable phase $\delta_2-\delta_0$ between $A_0$ and $A_2$
  
The various relations advocated in \cite{KDHMW} follow if
one sets 
\be
\label{constraint1}
\tilde K_1 = \tilde K_4 = 0\quad\mbox{and}\quad
\tilde K_8 = \cdots = \tilde K_{11} = 0\,.
\ee
The resulting fit values are in Table \ref{tab:fitKi}, column 2.
Several of the $\tilde K_i$ are uncomfortably large but inspection of the
size of the $p^2$ versus $p^4$ contributions shows nothing conspicuous.
The source of the problem turned out to be $\gamma_3$ which with the
constraints of Eq. (\ref{constraint1}) can only be fitted by varying
$\tilde K_6$ but it comes multiplied with a small factor. It then
produces fairly large other $\tilde K_i$ in order to minimize
the effect of $\tilde K_7$ on the other quantities.
We have therefore also fitted using the constraints
\be
\label{constraint2}
\tilde K_1 = \tilde K_4 = \tilde K_6 = 0\quad\mbox{and}\quad
\tilde K_8 = \cdots = \tilde K_{11} = 0\,.
\ee

We can now attempt to also determine somewhat the combinations
$\tilde K_1$ and $\tilde K_4$. Leaving $\tilde K_1$ free hardly changes
the quality of the fit but shows that the actual value of $G_8$
can change over a fairly wide range. The results are shown in the fourth column
of Table \ref{tab:fitKi} with constraints
\be
\label{constraint3}
\tilde K_4 = 0\quad\mbox{and}\quad
\tilde K_8 = \cdots = \tilde K_{11} = 0\,.
\ee
The same type of correlation occurs for $G_{27}$ and $\tilde K_4$ as shown
with the results in column 5 of the Table where we fitted with
constraints
\be
\label{constraint4}
\tilde K_1 = 0,\quad \tilde K_4=0.1\quad\mbox{and}\quad
\tilde K_8 = \cdots = \tilde K_{11} = 0\,.
\ee
The reason we have picked a fixed value for $\tilde K_4$
is because there is very long, narrow and extremely shallow
fitting region that moves in the end to a very small $G_{27}$ and values
for $\tilde K_4$, $\tilde K_5$, $\tilde K_6$ and $\tilde K_7$ that
are simply enormous but a total $\chi^2$ which is only marginally smaller
than the one shown, 11.7 rather than 11.9.

\begin{table}
\begin{center}
\begin{tabular}{|c|c|c|c|c|c|}
\hline
constraints & Eq. (\ref{constraint1}) &Eq. (\ref{constraint2}) &
Eq. (\ref{constraint3}) &Eq. (\ref{constraint4})& Eq. (\ref{constraint1}) \\
\hline
Fitted      &\multicolumn{4}{c|}{$|A_0|,\ldots,\xi_3^\prime$}& experiment   \\
\hline
$G_8$          &  5.47(2)          & 5.47(2)    &7.24(2.04)     &5.49&5.45(2)\\
$G_{27}$       &  0.392(2)         &0.392(2)    &0.392(2)   &0.139 &0.392(2)\\
$10^3\tilde K_{1}/G_8$ & $\equiv0$&$\equiv0$ &$-$8.5(7.5)&$\equiv0$&$\equiv0$\\
$10^3\tilde K_{2}/G_8$ & 54.7(2.8)&53.6(2.7)   &41.3(11.4)    &53.7&51.9(3.2)\\
$10^3\tilde K_{3}/G_8$ & 3.0(1.4) &3.5(1.3)    &10.0(6.3)   &3.2   &3.8(1.5)\\
$10^3\tilde K_{4}/G_{27}$&$\equiv0$&$\equiv0$ &$\equiv0$&$\equiv100$&$\equiv0$\\
$10^3\tilde K_{5}/G_{27}$ &$-$54.5(23.4)&$-$19.8(9.6)&$-$54.5(23.1)&$-$49.4&$-$42.5(16.6)\\
$10^3\tilde K_{6}/G_{27}$ &$-$185(114)&$\equiv0$&$-$184(113)&$-$366&$-$166(113)\\
$10^3\tilde K_{7}/G_{27}$ &114(46)&45.1(18.2) & 114(46)   &199&120(32) \\
\hline
$\chi^2/DOF$    & 12.3/5        &14.9/6      &11.8/4          &11.9/4&26.8/10\\
\hline
\end{tabular}
\end{center}
\caption{The results for $G_8$, $G_{27}$ and the $\tilde K_i$
for the various constraints
discussed in the text. In brackets are the MINUIT errors.}
\label{tab:fitKi}
\end{table}

How well do the relations proposed in \cite{KDHMW} stand up to the new data.
In practice, for the octet which dominates the $\Delta I=1/2$ part,
it means that $\alpha_1$ determines $\zeta_1$,
$\beta_1$ determines $\xi_1$. The first of these relations is
extremely well satisfied, but it is hard to get $\xi_1$ to better
than about 2.5 standard deviations, the value obtained from the
various above fits is
\be
\label{predxi}
\xi_1 \approx (-0.98-i0.13)\cdot10^{-8}\,.
\ee
Inspection of the numerical coefficients for the octet
parameters that only contribute suppressed by powers of $m_\pi^2$
reveals that some of them show up with large numerical coefficients.
One can then instead choose to leave $K_{8}$ or $\tilde K_9$
free rather than $\tilde K_1$. The resulting fits are of similar quality
to the one with $\tilde K_1$ free and do not significantly improve
the prediction for $\xi_1$ given in (\ref{predxi}).
The possible variation is less than 10\%.

There is no similar single value with significant discrepancies
for the $\Delta I = 3/2$ parameters. The above named problem with $\gamma_3$
does not lead to a major discrepancy but typically the
quadratic quantities are the ones with a deviation of about
0.5 to 1.4 $\sigma$.

Varying the masses of the pion and kaon or the eta mass used in the loops
does not change the results of the fits significantly.

The fit via the intermediate step of $|A_0|,\ldots,\xi_3^\prime$
is fairly fast and allows an easy study of the variation with the inputs.
We have also done the fit directly to all the experimental data listed
in Tables \ref{tab:widths} and \ref{tab:dalitz} with only the
use of Re($\lambda$) for the $K_S\to\pi^+\pi^-\pi^0$ data.
The masses used in the phase space were the physical masses occurring
in the decays. The masses used in the amplitudes are the physical
kaon mass and the pion mass such that $3m_\pi^2=
\sum_{i=1,3} m_{\pi^i}^2$ is satisfied.
The eta mass in the loops we then calculated using the GMO relation.
The resulting values for the $\tilde K_i$ are given in the last column
of Table \ref{tab:fitKi} and the values for the widths and
Dalitz plot distribution variables are in the column labeled
CHPT fit in Tables \ref{tab:widths} and \ref{tab:dalitz}. We have
allowed an extra phase between $A_0$ and $A_2$ in this fit.
Otherwise this fit corresponds to the constraints (\ref{constraint1}).
The higher $\chi^2$ is mainly due to the fact that $\xi_1$ has been
fitted to more than one quantity, and here we have the
discrepancy with the slope $k$ in $K_L\to\pi^+\pi^-\pi^0$
and $K^+\to\pi^0\pi^0\pi^+$. These together account for 16.4 of $\chi^2$.

The fact that the last fit and the fit via the intermediate step
of $|A_0|,\ldots,\xi_3^\prime$ agree very well is an indication that the
isospin breaking effects at the amplitude level are small. An estimate
can be done by calculating the amplitudes and the slopes
with the masses of ${K^0}$, ${\pi^0}$ and comparing them with the
amplitudes calculated with $m_{K^+}$, $m_{\pi^+}$. Since often factors
of $m_K^2-m_\pi^2$ appear this is the worst case. The changes
are typically 2\% for the amplitudes, 5-8\% for the linear slopes
and up to 20\% for the quadratic slopes.

\section{Conclusions}
\label{conclusions}

In this paper we have recalculated the $K\to\pi\pi$ and $K\to\pi\pi\pi$
amplitudes to next-to-leading order in CHPT in the isospin limit
and presented the amplitudes in a simple form suitable for
dispersive estimates of higher orders.

We have performed a new global fit to all available kaon data
at this level taking into account the isospin breaking in the
phase space. Work is under progress to include isospin breaking
also in the amplitudes.

At present the situation is that a satisfactory agreement is obtained
with the predictions keeping in mind that for the quadratic slopes
higher order corrections could be sizable as happened with the analogous
results in $\eta\to3\pi$ results. A discussion of the latter
together with references can be found in \cite{BG}. Some work
exists on dispersive corrections in $K\to3\pi$ \cite{DAmbrosio},
but more study is needed to fit their results in this framework.

Now the isospin breaking corrections, including possible
electromagnetic radiative corrections,
need to be studied to see if they are the cause of the differences
in the quadratic Dalitz plot parameters. Work exists to a large extent for
the $K\to2\pi$ decays \cite{iso}, but little has been done using
modern methods for $K\to3\pi$ decays.

\section*{Acknowledgments}
This work is supported by the Swedish Research Council
 and
the European Union TMR network, Contract No. ERB\-FMRX--CT980169 
(EURODAPHNE).

\appendix
\renewcommand{\theequation}{\Alph{section}.\arabic{equation}}
\section{Ambiguity in the definition of $M_i$}
\setcounter{equation}{0}
\label{App:ambiguity}

The functions $M_i(s)$ defined in Eq. (\ref{defMi}) are not unique
for two reasons. The variables $s_1, s_2$ and $s_3$ are not
independent, they satisfy the relation (\ref{defsi}),
and low powers of $s_1,s_2$ and $s_3$ can be fitted in more than one
of the $M_i$ functions.

As a simple example, we could have added
\be
\delta M_0(s) = \alpha \left(s -s_0\right)\,,
\ee
with $\alpha$ any constant, without changing the result,
for $A^L_{000}$.
The full set of these ambiguities can be derived by checking
the total number of independent terms that exist in the
polynomial expansion of the amplitudes and comparing it with the
the same expansion of the $M_i(s)$.

The changes of the functions that leave all amplitudes unchanged are
\ba
M_0(s) &\longrightarrow& M_0(s) + \delta_{1} \left(s-s_0\right)
\\ 
\label{ambiM2}
M_1(s) &\longrightarrow& M_1(s) + \delta_{2},
\quad
M_2(s) \longrightarrow M_2(s) -\frac{1}{2} \delta_{2}\,;
\\
M_1(s) &\longrightarrow& M_1(s) + \delta_{3}\left(s-s_0)\right),
\quad
M_2(s) \longrightarrow M_2(s) + \delta_{3}\left(s-s_0)\right)\,;
\\
M_1(s) &\longrightarrow&M_1(s)+2\delta_4 s,
\quad
M_2(s) \longrightarrow M_2(s)-\delta_4 s,
\quad
M_3(s) \longrightarrow M_3(s)+\delta_4\,;
\\
M_1(s) &\longrightarrow&M_1(s)-2\delta_5 s\left( s-3 s_0\right),
\quad
M_2(s) \longrightarrow M_2(s)+\delta_5   s\left( s-3 s_0\right),
\nonumber\\
M_3(s) &\longrightarrow& M_3(s)+\delta_5 s\,;
\\
\label{ambiM6}
M_1(s) &\longrightarrow&M_1(s)+\frac{2}{3}\delta_6 \left( s-3 s_0\right)^3,
\quad
M_2(s) \longrightarrow M_2(s)-\frac{1}{3}\delta_6   s^2\left( s-9 s_0\right),
\nonumber\\
M_3(s) &\longrightarrow& M_3(s)+\delta_6 s^2\,;
\\
M_4(s) &\longrightarrow& M_4(s)+\delta_7\,;
\\
M_4(s) &\longrightarrow& M_4(s)+\delta_8 s\,,
\quad
M_5(s) \longrightarrow M_5(s)+\delta_8;
\\
M_5(s) &\longrightarrow& M_5(s)+\delta_9,
\quad
M_6(s) \longrightarrow M_5(s)+\delta_9;
\\
M_4(s) &\longrightarrow& M_4(s)-\delta_{10} s\left( s-3 s_0\right)\,,
\quad
M_5(s) \longrightarrow M_5(s)+\delta_{10}s;
\\
M_5(s) &\longrightarrow& M_5(s)+\delta_{11}s,
\quad
M_6(s) \longrightarrow M_5(s)+\delta_{11}s;
\\
M_4(s) &\longrightarrow& M_4(s)+\delta_{12}\left(s^3-9s_0s^2+18s_0^2s\right),
\quad
M_5(s) \longrightarrow M_5(s)+\delta_{12}s^2;
\nonumber\\
M_6(s) &\longrightarrow& M_6(s)-2\delta_{12}s^2;
\\
M_4(s) &\longrightarrow& M_4(s)+\delta_{13}3s_0
\left(s^3-9s_0s^2+18s_0^2s\right),
\quad
M_5(s) \longrightarrow M_5(s)+\delta_{13}s^3,
\nonumber\\
M_6(s) &\longrightarrow& M_6(s)-2\delta_{13}\left(s^2-9s_0s^2\right)\,.
\ea
The triplets $(M_7,M_8,M_9)$ and $(M_{10},M_{11},M_{12})$ have an ambiguity
similar to (\ref{ambiM2})-(\ref{ambiM6}), leading to a total ambiguity
of 23 free parameters.

\section{Explicit expressions for the $M_i(s)$ in $K\to3\pi$.}
\label{App:results}
\setcounter{equation}{0}

We write the $M_i(s)$ functions defined in Eq. (\ref{defMi})
as
\be
M_i(s) = \left.M_i(s)\right|_{p^2}
+i\frac{CF_0^4}{F_\pi^3 F_K}
\left\{\frac{G_8}{F_\pi^2}M_i^8(s)+\frac{G_8^\prime}{F_\pi^2}M_i^{8\prime}(s)
+\frac{G_{27}}{F_\pi^2}M_i^{27}(s)\right\}\,.
\ee

The effect of $G_8^\prime$ cannot be distinguished from higher order
coefficients in decays not involving external fields. This was known at tree
level earlier and has been proven to one-loop in \cite{KMW1}.
The effect of
$M_i^{8\prime}(s)$ can be reconstructed from $M_i^8(s)$ by the changes
\ba
N_{10}^r+2 N_{11}^r+N_{12}^r &\longrightarrow&
N_{10}^r+2 N_{11}^r+N_{12}^r +\frac{G_8^\prime}{G_8}
\left(2L_5^r-16L_6^r-8L_8^r\right)\,,
\nonumber\\
N_{7}^r &\longrightarrow&
N_{7}^r+\frac{G_8^\prime}{G_8}
\left(-4L_4^r \right)\,.
\ea

The results for the $K\to\pi\pi$ amplitudes are in full agreement
with \cite{BPP} and can be found there.

We have brought the $M_i^j(s)$ in a form that satisfies the
isospin relations (\ref{isorelations}). These can be used in the form
\ba
M_1^j(s)&=&M_0^j(s)-2M_2^j(s)\,,
\nonumber\\
M_6^j(s)&=&M_5^j(s)-M_9^j(s)-M_{12}^j(s)\,,
\nonumber\\
M_7^j(s)&=&2M_4^j(s)-\frac{1}{2}M_{10}^j(s)+M_{11}^j(s)\,,
\nonumber\\
M_8^j(s)&=&-M_4^j(s)+\frac{1}{2}M_{10}^j(s)
\ea
to reconstruct these.
We have extensively used the GMO relation in writing them in this form.

The explicit expressions
for  $\overline A$, $\overline B$
and $\overline B_1$, the finite part of the loop functions, can be found
in many places, e.g. \cite{ABT1}.

The octet ones are:
\ba
  \lefteqn{ M^8_0(s) =}&&
\nonumber\\&&
       + (N_1^r+N_2^r)\, (  - 2\,m_\pi^2\,m_K^2 + 3\,m_\pi^2\,s - 2\,m_\pi^4 + m_K^2\,s - s^2 )
\nonumber\\&&
       + N_3^r \, ( 2\,m_\pi^2\,m_K^2 - 3\,m_\pi^2\,s + 5\,m_\pi^4 - m_K^2\,s + m_K^4 - 2\,s^2 )
\nonumber\\&&
       + N_5^r \, ( 2\,m_\pi^2\,m_K^2 - 2\,m_\pi^2\,s + 2\,m_\pi^4 + 2\,m_K^2\,s )
%\nonumber\\&&
       + N_7^r \, ( 8\,m_\pi^2\,m_K^2 - 4\,m_K^2\,s )
\nonumber\\&&
       + N_8^r \, (  - 2\,m_\pi^2\,s + 2\,m_\pi^4 + 4\,m_K^2\,s )
%\nonumber\\&&
       + N_9^r \, (  - 2\,m_\pi^2\,m_K^2 - 2\,m_\pi^2\,s + 2\,m_\pi^4 + 2\,m_K^2\,s )
\nonumber\\&&
       + (N_{10}^r+2N_{11}^r+N_{12}^r) \, (  - 4\,m_\pi^2\,m_K^2 )
\nonumber\\&&
       + L_1^r \, ( 32\,m_\pi^2\,m_K^2 - 48\,m_\pi^2\,s + 32\,m_\pi^4 - 16\,m_K^2\,s + 16\,s^2 )
\nonumber\\&&
       + L_2^r \, (  - 16\,m_\pi^2\,m_K^2 + 24\,m_\pi^2\,s - 40\,m_\pi^4 + 8\,m_K^2\,s - 8\,m_K^4 + 
         16\,s^2 )
\nonumber\\&&
       + L_3^r \, ( 16\,m_\pi^2\,m_K^2 - 24\,m_\pi^2\,s + 16\,m_\pi^4 - 8\,m_K^2\,s + 8\,s^2 )
\nonumber\\&&
       + L_4^r \, (  - 16\,m_\pi^2\,m_K^2 + 16\,m_\pi^2\,s - 16\,m_\pi^4 )
%\nonumber\\&&
       + L_5^r \, (  - 8\,m_\pi^2\,m_K^2 + 16\,m_\pi^2\,s - 16\,m_\pi^4 )
\nonumber\\&&
       + ({1}/{(16\pi^2)}) \, ( 4/3\,m_\pi^2\,m_K^2 + 1/2\,m_\pi^2\,s - 1/2\,m_\pi^4 - 3/2\,m_K^2\,s + 1/2\,
         m_K^4 )
\nonumber\\&&
       + \overline{A}(m_\pi^2) \, (  - 19/8\,m_\pi^2 - 2/3\,m_K^2 + 21/8\,s )
%\nonumber\\&&
       + \overline{A}(m_K^2) \, (  - 13/4\,m_\pi^2 - 1/3\,m_K^2 + 3\,s )
\nonumber\\&&
       + \overline{A}(m_\eta^2) \, (  - 3/8\,m_\pi^2 + 3/8\,s )
%\nonumber\\&&
       + \overline{B}(m_\pi^2,m_\pi^2,s) \, ( 1/2\,m_\pi^2\,m_K^2 - 2\,m_\pi^2\,s + m_\pi^4 + s^2 )
\nonumber\\&&
       + \overline{B}(m_\pi^2,m_K^2,s) \, ( 13/8\,m_\pi^2\,m_K^2 - 5/4\,m_\pi^2\,s - 5/8\,m_K^2\,s - 1/8\,m_K^4
          + 3/4\,s^2 )
\nonumber\\&&
       + \overline{B}(m_K^2,m_K^2,s) \, ( 1/4\,m_\pi^2\,s + 1/4\,m_K^2\,s - 1/4\,s^2 )
\nonumber\\&&
       + \overline{B}(m_\eta^2,m_K^2,s) \, (  - 5/24\,m_\pi^2\,m_K^2 + 1/8\,m_K^2\,s + 1/24\,m_K^4 )
%\nonumber\\&&
       + \overline{B}(m_\eta^2,m_\eta^2,s) \, ( 1/18\,m_\pi^2\,m_K^2 )
\nonumber\\&&
       + \overline{B_1}(m_\pi^2,m_K^2,s) \, (  - 3/4\,m_\pi^2\,m_K^2 + 1/2\,m_\pi^4 + 1/4\,m_K^4 )
\nonumber\\&&
       + \overline{B_1}(m_\eta^2,m_K^2,s) \, ( 1/4\,m_\pi^2\,m_K^2 - 1/4\,m_K^4 )\,.
\ea
\ba
 \lefteqn{  M^8_2(s) =}&&
\nonumber\\&&
       + N_3^r \, (  - s^2 )
%\nonumber\\&&
       + L_2^r \, ( 8\,s^2 )
%\nonumber\\&&
       + ({1}/{(16\pi^2)}) \, ( 1/6\,s^2 )
\nonumber\\&&
       + \overline{B}(m_\pi^2,m_\pi^2,s) \, ( 1/2\,m_\pi^2\,m_K^2 - 3/4\,m_\pi^2\,s + 1/2\,m_\pi^4 - 1/4\,m_K^2\,s
          + 1/4\,s^2 )
\nonumber\\&&
       + \overline{B}(m_\pi^2,m_K^2,s) \, ( 1/2\,m_\pi^2\,m_K^2 - 1/2\,m_\pi^2\,s + 1/4\,m_\pi^4 - 1/2\,m_K^2\,s
          + 1/4\,m_K^4 + 1/4\,s^2 ).
\nonumber\\
\ea
\ba
\lefteqn{M^8_3(s) =}&&
\nonumber\\&&
       + \overline{B}(m_\pi^2,m_\pi^2,s) \, ( 1/3\,m_\pi^2 - 1/12\,s )
%\nonumber\\&&
       + \overline{B}(m_\pi^2,m_K^2,s) \, ( 1/8\,m_\pi^2 + 1/24\,m_K^2 - 1/24\,s )
\nonumber\\&&
       + \overline{B}(m_\eta^2,m_K^2,s) \, (  - 1/8\,m_\pi^2 + 5/8\,m_K^2 - 1/8\,s )
%\nonumber\\&&
       + \overline{B_1}(m_\pi^2,m_K^2,s) \, (  - 1/12\,m_\pi^2 + 1/12\,m_K^2 )
\nonumber\\&&
       + \overline{B_1}(m_\eta^2,m_K^2,s) \, ( 1/12\,m_\pi^2 - 1/12\,m_K^2 ).
\ea
\ba
   M^8_4(s) &=& 0.
\ea
\ba
\lefteqn{   M^8_5(s) =}&&
\nonumber\\&&
       + \overline{B}(m_\pi^2,m_K^2,s) \, (  - 1/8\,m_\pi^2 - 1/24\,m_K^2 + 1/24\,s )
%\nonumber\\&&
       + \overline{B}(m_K^2,m_K^2,s) \, (  - 1/3\,m_K^2 + 1/12\,s )
\nonumber\\&&
       + \overline{B}(m_\eta^2,m_K^2,s) \, ( 1/8\,m_\pi^2 - 5/8\,m_K^2 + 1/8\,s )
%\nonumber\\&&
       + \overline{B_1}(m_\pi^2,m_K^2,s) \, ( 1/12\,m_\pi^2 - 1/12\,m_K^2 )
\nonumber\\&&
       + \overline{B_1}(m_\eta^2,m_K^2,s) \, (  - 1/12\,m_\pi^2 + 1/12\,m_K^2 );
\ea
\ba
 M^8_9(s) &=& -M^8_3(s).
\ea
\ba
\lefteqn{   M^8_{10}(s) =}&&
\nonumber\\&&
       + (N_1^r+N_2^r) \, (  - 2\,m_\pi^2\,m_K^2 + 3\,m_\pi^2\,s - 5\,m_\pi^4 + m_K^2\,s - m_K^4 )
\nonumber\\&&
       + N_3^r \, ( 2\,m_\pi^2\,m_K^2 - 3\,m_\pi^2\,s - m_\pi^4 - m_K^2\,s - m_K^4 + 2\,s^2 )
\nonumber\\&&
       + N_5^r \, (  - 8\,m_\pi^2\,m_K^2 - 2\,m_\pi^2\,s + 2\,m_\pi^4 + 2\,m_K^2\,s - 2\,m_K^4 )
\nonumber\\&&
       + N_7^r \, (  - 4\,m_\pi^2\,m_K^2 - 4\,m_K^2\,s + 4\,m_K^4 )
\nonumber\\&&
       + N_8^r \, (  - 10\,m_\pi^2\,m_K^2 - 2\,m_\pi^2\,s + 2\,m_\pi^4 + 4\,m_K^2\,s - 4\,m_K^4 )
\nonumber\\&&
       + N_9^r \, (  - 2\,m_\pi^2\,s + 2\,m_\pi^4 + 2\,m_K^2\,s - 2\,m_K^4 )
%\nonumber\\&&
       + (N_{10}^r+2N_{11}^r+N_{12}^r) \, ( 8\,m_\pi^2\,m_K^2 )
\nonumber\\&&
       + L_1^r \, ( 32\,m_\pi^2\,m_K^2 - 48\,m_\pi^2\,s + 80\,m_\pi^4 - 16\,m_K^2\,s + 16\,m_K^4 )
\nonumber\\&&
       + L_2^r \, (  - 16\,m_\pi^2\,m_K^2 + 24\,m_\pi^2\,s + 8\,m_\pi^4 + 8\,m_K^2\,s + 8\,m_K^4 - 16
         \,s^2 )
\nonumber\\&&
       + L_3^r \, ( 16\,m_\pi^2\,m_K^2 - 24\,m_\pi^2\,s + 40\,m_\pi^4 - 8\,m_K^2\,s + 8\,m_K^4 )
\nonumber\\&&
       + L_4^r \, ( 16\,m_\pi^2\,m_K^2 + 16\,m_\pi^2\,s - 16\,m_\pi^4 )
%\nonumber\\&&
       + L_5^r \, ( 16\,m_\pi^2\,s - 16\,m_\pi^4 )
\nonumber\\&&
       + ({1}/{(16\pi^2)}) \, ( 11/9\,m_\pi^2\,m_K^2 + 4/9\,m_\pi^2\,s - 1/3\,m_\pi^4 - 13/9\,m_K^2\,s + 4/9\,
         m_K^4 - 1/3\,s^2 )
\nonumber\\&&
       + \overline{A}(m_\pi^2) \, (  - 25/8\,m_\pi^2 - 31/24\,m_K^2 + 21/8\,s )
%\nonumber\\&&
       + \overline{A}(m_K^2) \, (  - 5/2\,m_\pi^2 - 7/3\,m_K^2 + 3\,s )
\nonumber\\&&
       + \overline{A}(m_\eta^2) \, (  - 3/8\,m_\pi^2 - 3/8\,m_K^2 + 3/8\,s )
\nonumber\\&&
       + \overline{B}(m_\pi^2,m_\pi^2,s) \, (  - m_\pi^2\,m_K^2 + 3/2\,m_\pi^2\,s - m_\pi^4 + 1/2\,m_K^2\,s - 1/2
         \,s^2 )
\nonumber\\&&
       + \overline{B}(m_\pi^2,m_K^2,s) \, (  - m_\pi^2\,m_K^2 + m_\pi^2\,s - 1/2\,m_\pi^4 + m_K^2\,s - 1/2\,
         m_K^4 - 1/2\,s^2 ).
\ea
\ba
\lefteqn{   M^8_{11}(s) =}&&
\nonumber\\&&
       +( N_1^r + N_2^r + N_3^r- 16 L_1^r  - 8 L_2^r - 8 L_3^r) \, ( s^2 )
\nonumber\\&&
       + ({1}/{(16\pi^2)}) \, (  - 1/18\,m_\pi^2\,s + 1/18\,m_K^2\,s + 1/6\,s^2 )
\nonumber\\&&
       + \overline{B}(m_\pi^2,m_\pi^2,s) \, ( 5/4\,m_\pi^2\,s - 1/2\,m_\pi^4 - 1/4\,m_K^2\,s - 3/4\,s^2 )
\nonumber\\&&
       + \overline{B}(m_\pi^2,m_K^2,s) \, (  - 9/8\,m_\pi^2\,m_K^2 + 3/4\,m_\pi^2\,s + 1/4\,m_\pi^4 + 1/8\,m_K^2\,
         s + 3/8\,m_K^4 - 1/2\,s^2 )
\nonumber\\&&
       + \overline{B}(m_K^2,m_K^2,s) \, (  - 1/4\,m_\pi^2\,s - 1/4\,m_K^2\,s + 1/4\,s^2 )
\nonumber\\&&
       + \overline{B}(m_\eta^2,m_K^2,s) \, ( 5/24\,m_\pi^2\,m_K^2 - 1/8\,m_K^2\,s - 1/24\,m_K^4 )
%\nonumber\\&&
       + \overline{B}(m_\eta^2,m_\eta^2,s) \, (  - 1/18\,m_\pi^2\,m_K^2 )
\nonumber\\&&
       + \overline{B_1}(m_\pi^2,m_K^2,s) \, ( 3/4\,m_\pi^2\,m_K^2 - 1/2\,m_\pi^4 - 1/4\,m_K^4 )
\nonumber\\&&
       + \overline{B_1}(m_\eta^2,m_K^2,s) \, (  - 1/4\,m_\pi^2\,m_K^2 + 1/4\,m_K^4 ).
\ea
\ba
 M^8_{12}(s) &=& M^8_3(s) = -M^8_9(s).
\ea

The 27 plet ones depend as well on the quantity
\be
P_{K\pi} = \frac{1}{m_K^2-m_\pi^2}\,.
\ee
\ba
\lefteqn{   M^{27}_0(s) =}&&
\nonumber\\&&
       + D_1^r \, (  - 4/3\,m_\pi^2\,m_K^2 )
\nonumber\\&&
       + D_2^r \, (( 2/3\,m_\pi^2\,m_K^2 + 10\,m_\pi^2\,s - 40/3\,m_\pi^4 )
%\nonumber\\&&
       + P_{K\pi}
%D_2^r \,
 ( 10\,m_\pi^4\,s - 40/3\,m_\pi^6 ))
\nonumber\\&&
       + D_4^r \,( (  - 11/2\,m_\pi^2\,m_K^2 + m_\pi^2\,s + 7/3\,m_\pi^4 + 3/2\,m_K^2\,s - 5/6\,m_K^4
          )
\nonumber\\&&
~~       + P_{K\pi}\,
%D_4^r \,
 (  - 10\,m_\pi^4\,s + 40/3\,m_\pi^6 ))
\nonumber\\&&
       + D_5^r \, ( 29/6\,m_\pi^2\,m_K^2 - 11\,m_\pi^2\,s + 11\,m_\pi^4 - 3/2\,m_K^2\,s + 5/6\,m_K^4
          )
\nonumber\\&&
       + D_6^r \, ( 5/3\,m_\pi^2\,m_K^2 + 1/3\,m_\pi^2\,s - 1/3\,m_\pi^4 - m_K^2\,s )
\nonumber\\&&
       + D_7^r \, ((  - 25/3\,m_\pi^2\,m_K^2 - 3\,m_\pi^2\,s + 8\,m_\pi^4 + 6\,m_K^2\,s - 10/3\,m_K^4
          )
\nonumber\\&&
~~       + P_{K\pi}\,
D_7^r \,
 (  - 15\,m_\pi^4\,s + 20\,m_\pi^6 ))
\nonumber\\&&
       + D_{26}^r \, (  - 2/3\,m_\pi^2\,m_K^2 + m_\pi^2\,s - 2/3\,m_\pi^4 + 1/3\,m_K^2\,s - 1/3\,s^2
          )
\nonumber\\&&
       + D_{27}^r \, ( 4/3\,m_\pi^2\,m_K^2 - 2\,m_\pi^2\,s + 10/3\,m_\pi^4 - 2/3\,m_K^2\,s + 2/3\,m_K^4
          - 4/3\,s^2 )
\nonumber\\&&
       + D_{28}^r \, ( 8\,m_\pi^2\,m_K^2 - 12\,m_\pi^2\,s + 12\,m_\pi^4 - 4\,m_K^2\,s + 4/3\,m_K^4 )
\nonumber\\&&
       + D_{29}^r \, (  - 3/2\,m_\pi^2\,m_K^2 + 9/4\,m_\pi^2\,s - 13/4\,m_\pi^4 + 3/4\,m_K^2\,s - 7/12
         \,m_K^4 + s^2 )
\nonumber\\&&
       + D_{30}^r \, ( 3/2\,m_\pi^2\,m_K^2 - 9/4\,m_\pi^2\,s + 1/4\,m_\pi^4 - 3/4\,m_K^2\,s - 5/12\,
         m_K^4 + 2\,s^2 )
\nonumber\\&&
       + D_{31}^r \, ( 4\,m_\pi^2\,m_K^2 - 6\,m_\pi^2\,s + 4\,m_\pi^4 - 2\,m_K^2\,s + 2\,s^2 )
\nonumber\\&&
       + L_1^r \, (  - 32\,m_\pi^2\,m_K^2 + 48\,m_\pi^2\,s - 32\,m_\pi^4 + 16\,m_K^2\,s - 16\,s^2 )
\nonumber\\&&
       + L_2^r \, ( 16\,m_\pi^2\,m_K^2 - 24\,m_\pi^2\,s + 40\,m_\pi^4 - 8\,m_K^2\,s + 8\,m_K^4 - 16\,
         s^2 )
\nonumber\\&&
       + L_3^r \, ((  - 58/3\,m_\pi^2\,m_K^2 + 34\,m_\pi^2\,s - 118/3\,m_\pi^4 + 8\,m_K^2\,s - 8\,s^2
          )
\nonumber\\&&
~~       + P_{K\pi}\,
%L_3^r \,
 ( 40\,m_\pi^4\,s - 160/3\,m_\pi^6 ))
\nonumber\\&&
       + L_4^r \, ( 16\,m_\pi^2\,m_K^2 - 16\,m_\pi^2\,s + 16\,m_\pi^4 )
\nonumber\\&&
       + L_5^r \, (( 8\,m_\pi^2\,m_K^2 - 16\,m_\pi^2\,s + 68/3\,m_\pi^4 )
%\nonumber\\&&
       + P_{K\pi}\,
%L_5^r \,
 (  - 20\,m_\pi^4\,s + 80/3\,m_\pi^6 ))
\nonumber\\&&
       + ({1}/{(16\pi^2)}) \,( ( 503/324\,m_\pi^2\,m_K^2 + 197/216\,m_\pi^2\,s - 287/216\,m_\pi^4 
 - 401/216\,    m_K^2\,s
\nonumber\\&&
~~ + 401/648\,m_K^4 )
%\nonumber\\&&
       + P_{K\pi}\,
({1}/{(16\pi^2)}) \,
 ( 5/4\,m_\pi^4\,s - 5/3\,m_\pi^6 ))
\nonumber\\&&
       + \overline{A}(m_\pi^2) \,( (  - 409/72\,m_\pi^2 - 49/9\,m_K^2 + 29/3\,s )
%\nonumber\\&&
       + P_{K\pi}\,
%\overline{A}(m_\pi^2) \,
 (  - 215/24\,m_\pi^2\,s + 215/18\,m_\pi^4 )
\nonumber\\&&
~~       + P_{K\pi}^2\,
%\overline{A}(m_\pi^2) \,
 ( 5/2\,m_\pi^4\,s - 10/3\,m_\pi^6 ))
\nonumber\\&&
       + \overline{A}(m_K^2) \, ((  - 191/72\,m_\pi^2 - 19/8\,m_K^2 + 37/6\,s )
%\nonumber\\&&
       + P_{K\pi}\,
%\overline{A}(m_K^2) \,
 (  - 325/24\,m_\pi^2\,s + 325/18\,m_\pi^4 )
\nonumber\\&&
~~       + P_{K\pi}^2\,
%\overline{A}(m_K^2) \,
 (  - 5/2\,m_\pi^4\,s + 10/3\,m_\pi^6 ))
\nonumber\\&&
       + \overline{A}(m_\eta^2) \, ((  - 13/12\,m_\pi^2 - 5/8\,m_K^2 + 3/2\,s )
%\nonumber\\&&
       + P_{K\pi}\,
%\overline{A}(m_\eta^2) \,
 (  - 5/4\,m_\pi^2\,s + 5/3\,m_\pi^4 ))
\nonumber\\&&
       + \overline{B}(m_\pi^2,m_\pi^2,s) \,( ( 1/3\,m_\pi^2\,m_K^2 - 13/6\,m_\pi^2\,s + 2/3\,m_\pi^4 - 5/6\,m_K^2\,s
          + 3/2\,s^2 )
\nonumber\\&&
~~       + P_{K\pi}\,
%\overline{B}(m_\pi^2,m_\pi^2,s) \,
 (  - 5/2\,m_\pi^2\,s^2 + 35/6\,m_\pi^4\,s - 10/
         3\,m_\pi^6 ))
\nonumber\\&&
       + \overline{B}(m_\pi^2,m_K^2,s) \,( ( 91/24\,m_\pi^2\,m_K^2 - 35/6\,m_\pi^2\,s + 10/3\,m_\pi^4 
  - 25/12\,
         m_K^2\,s - 7/24\,m_K^4
\nonumber\\&&
~~ + 19/8\,s^2 )
%\nonumber\\&&
       + P_{K\pi}\,
%\overline{B}(m_\pi^2,m_K^2,s) \,
 ( 15/8\,m_\pi^2\,s^2 - 5\,m_\pi^4\,s + 10/3\,
         m_\pi^6 ))
\nonumber\\&&
       + \overline{B}(m_K^2,m_K^2,s) \, ((  - 13/12\,m_\pi^2\,s - 7/8\,m_K^2\,s + 7/8\,s^2 )
%\nonumber\\&&
       + P_{K\pi}\,
%\overline{B}(m_K^2,m_K^2,s) \,
 ( 5/8\,m_\pi^2\,s^2 - 5/6\,m_\pi^4\,s ))
\nonumber\\&&
       + \overline{B}(m_\eta^2,m_K^2,s) \, ( 5/24\,m_\pi^2\,m_K^2 - 1/8\,m_K^2\,s - 1/24\,m_K^4 )
%\nonumber\\&&
       + \overline{B}(m_\eta^2,m_\eta^2,s) \, (  - 1/18\,m_\pi^2\,m_K^2 )
\nonumber\\&&
       + \overline{B_1}(m_\pi^2,m_K^2,s) \, (  - 4/3\,m_\pi^2\,m_K^2 + 1/3\,m_\pi^4 - 2/3\,m_K^4 )
\nonumber\\&&
       + \overline{B_1}(m_\eta^2,m_K^2,s) \, (  - 1/4\,m_\pi^2\,m_K^2 + 1/4\,m_K^4 ).
\ea
\ba
\lefteqn{   M^{27}_2(s) =}&&
\nonumber\\&&
       + D_{27}^r \, (  - 2/3\,s^2 )
%\nonumber\\&&
       + D_{28}^r \, (  - 4/3\,s^2 )
%\nonumber\\&&
       + D_{29}^r \, ( 7/12\,s^2 )
%\nonumber\\&&
       + D_{30}^r \, ( 5/12\,s^2 )
%\nonumber\\&&
       + L_2^r \, (  - 8\,s^2 )
\nonumber\\&&
       + P_{K\pi}\,L_3^r \, ( 10/3\,m_\pi^2\,s^2 )
\nonumber\\&&
       + ({1}/{(16\pi^2)}) \,( (  - 55/648\,m_\pi^2\,s + 55/648\,m_K^2\,s + 8/27\,s^2 )
%\nonumber\\&&
       + P_{K\pi}\,
%({1}/{(16\pi^2)}) \,
 (  - 5/24\,m_\pi^2\,s^2 ))
\nonumber\\&&
       + \overline{B}(m_\pi^2,m_\pi^2,s) \,( (  - 1/12\,m_\pi^2\,m_K^2 - 11/12\,m_\pi^2\,s + 1/3\,m_\pi^4 + 1/24\,
         m_K^2\,s + 3/8\,s^2 )
\nonumber\\&&
~~       + P_{K\pi}\,
\overline{B}(m_\pi^2,m_\pi^2,s) \,
 (  - 5/8\,m_\pi^2\,s^2 + 25/12\,m_\pi^4\,s - 5/
         3\,m_\pi^6 ))
\nonumber\\&&
       + \overline{B}(m_\pi^2,m_K^2,s) \,( ( 13/4\,m_\pi^2\,m_K^2 - 67/16\,m_\pi^2\,s + 11/6\,m_\pi^4 
- 217/96\,m_K^2\,s + 1/6\,m_K^4 
\nonumber\\&&
~~ + 67/32\,s^2 )
%\nonumber\\&&
       + P_{K\pi}\,
\overline{B}(m_\pi^2,m_K^2,s) \,
 ( 25/32\,m_\pi^2\,s^2 - 5/3\,m_\pi^4\,s + 5/6\,
         m_\pi^6 ))
\nonumber\\&&
       + \overline{B}(m_\eta^2,m_K^2,s) \,( ( 5/72\,m_\pi^2\,m_K^2 - 5/12\,m_\pi^2\,s + 5/18\,m_\pi^4 
- 5/32\,m_K^2
         \,s - 5/72\,m_K^4 
\nonumber\\&&
~~+ 5/32\,s^2 )
%\nonumber\\&&
       + P_{K\pi}\,
%\overline{B}(m_\eta^2,m_K^2,s) \,
 ( 5/32\,m_\pi^2\,s^2 - 5/12\,m_\pi^4\,s + 5/18
         \,m_\pi^6 ))
\nonumber\\&&
       + \overline{B_1}(m_\pi^2,m_K^2,s) \, (  - 15/16\,m_\pi^2\,m_K^2 + 5/24\,m_\pi^4 - 25/48\,m_K^4 )
\nonumber\\&&
       + \overline{B_1}(m_\eta^2,m_K^2,s) \, ( 5/16\,m_\pi^2\,m_K^2 + 5/48\,m_K^4 ).
\ea
\ba
\lefteqn{   M^{27}_3(s) =}&&
\nonumber\\&&
       + \overline{B}(m_\pi^2,m_\pi^2,s) \, (( 1/2\,m_\pi^2 - 1/8\,s )
%\nonumber\\&&
       + P_{K\pi}\,
%\overline{B}(m_\pi^2,m_\pi^2,s) \,
 ( 5/24\,m_\pi^2\,s - 5/6\,m_\pi^4 ))
\nonumber\\&&
       + \overline{B}(m_\pi^2,m_K^2,s) \, ( 67/144\,m_\pi^2 + 43/288\,m_K^2 - 43/288\,s %)
%\nonumber\\&&
       + P_{K\pi}\,
%\overline{B}(m_\pi^2,m_K^2,s) \,
 (  - 5/288\,m_\pi^2\,s + 5/72\,m_\pi^4 ))
\nonumber\\&&
       + \overline{B}(m_K^2,m_K^2,s) \,( (  - 5/18\,m_\pi^2 + 5/9\,m_K^2 - 5/36\,s )
%\nonumber\\&&
       + P_{K\pi}\,
%\overline{B}(m_K^2,m_K^2,s) \,
 ( 5/72\,m_\pi^2\,s - 5/18\,m_\pi^4 ))
\nonumber\\&&
       + \overline{B}(m_\eta^2,m_K^2,s) \,( (  - 7/24\,m_\pi^2 + 5/32\,m_K^2 - 1/32\,s )
%\nonumber\\&&
   + P_{K\pi}\,
%\overline{B}(m_\eta^2,m_K^2,s) \,
  ( 5/96\,m_\pi^2\,s - 5/24\,m_\pi^4 ))
\nonumber\\&&
       + \overline{B_1}(m_\pi^2,m_K^2,s) \, (  - 19/72\,m_\pi^2 + 43/144\,m_K^2 )
%\nonumber\\&&
       + \overline{B_1}(m_\eta^2,m_K^2,s) \, ( 1/18\,m_\pi^2 - 1/48\,m_K^2 ).
\nonumber\\
\ea
\ba
\lefteqn{   M^{27}_4(s) =}&&
\nonumber\\&&
       + D_2^r \, ( 10/3\,m_\pi^2\,s  )
%\nonumber\\&&
       + P_{K\pi}\,
%D_2^r \, 
 ( 10/3\,m_\pi^4\,s ))
%\nonumber\\&&
       + D_4^r \,(  - 10\,m_\pi^2\,s  - 5/2\,m_K^2\,s )
%\nonumber\\&&
       + P_{K\pi}\,
%D_4^r \, 
(  - 10/3\,m_\pi^4\,s))
\nonumber\\&&
       + D_5^r \, (  20/3\,m_\pi^2\,s + 5/2\,m_K^2\,s )
%\nonumber\\&&
       + D_7^r \, ( ( - 25/3\,m_\pi^2\,s - 10\,m_K^2\,s )
%\nonumber\\&&
       + P_{K\pi}\,
%D_7^r \,
 (  - 5\,m_\pi^4\,s ))
\nonumber\\&&
       + D_{28}^r  ( 10\,m_\pi^2\,s + 10/3\,m_K^2\,s - 10/3\,s^2 )
%\nonumber\\&&
       + (D_{29}^r-D_{30}^r)  (  - 5/4\,m_\pi^2 s - 5/12\,m_K^2 s  + 5/12\,s^2 )
\nonumber\\&&
       + L_3^r \,( (  10/3\,m_\pi^2\,s )
%\nonumber\\&&
       + P_{K\pi}\,
%L_3^r \,
 (  - 10/3\,m_\pi^2\,s^2 + 40/3\,m_\pi^4\,s       ))
\nonumber\\&&
       + L_4^r \, ( 64/9\,m_\pi^2\,m_K^2 )
%\nonumber\\&&
       + L_5^r \,( P_{K\pi}\,
%L_5^r \,
 (  - 20/3\,m_\pi^4\,s ))
\nonumber\\&&
       + ({1}/{(16\pi^2)}) \, (  5/18\,m_\pi^2\,s + 205/108\,m_K^2\,
         s+ 55/54\,s^2 )
%\nonumber\\&&
       + P_{K\pi}\,
%({1}/{(16\pi^2)}) \,
 ( 5/24\,m_\pi^2\,s^2 + 5/12\,m_\pi^4\,s ))
\nonumber\\&&
       + \overline{A}(m_\pi^2) \,(( - 1105/72\,s )
%\nonumber\\&&
       + P_{K\pi}\,
%\overline{A}(m_\pi^2) \,
 (  - 95/24\,m_\pi^2\,s  )
%\nonumber\\&&
    + P_{K\pi}^2\,
%\overline{A}(m_\pi^2) \,
 ( 5/6\,m_\pi^4\,s  ))
\nonumber\\&&
       + \overline{A}(m_K^2) \,( ( - 335/36\,s )
%\nonumber\\&&
       + P_{K\pi}\,
%\overline{A}(m_K^2) \,
 (  - 85/24\,m_\pi^2\,s)
%\nonumber\\&&
       + P_{K\pi}^2\,
%\overline{A}(m_K^2) \, 
(  - 5/6\,m_\pi^4\,s  ))
\nonumber\\&&
       + \overline{A}(m_\eta^2) \,( ( - 35/24\,s )
%\nonumber\\&&
       + P_{K\pi}\,
%\overline{A}(m_\eta^2) \,
 (  - 5/12\,m_\pi^2\,s ))
\nonumber\\&&
       + \overline{B}(m_\pi^2,m_\pi^2,s) \,( ( 5/4\,m_\pi^2\,m_K^2 - 35/6\,m_\pi^2\,s + 25/6\,m_\pi^4 - 5/8\,m_K^2\,
         s + 15/8\,s^2 )
\nonumber\\&&
 ~~      + P_{K\pi}\,
%\overline{B}(m_\pi^2,m_\pi^2,s) \, 
( 5/8\,m_\pi^2\,s^2 - 25/12\,m_\pi^4\,s + 5/3\,
         m_\pi^6 ))
\nonumber\\&&
       + \overline{B}(m_\pi^2,m_K^2,s) \, ((  - 5/12\,m_\pi^2\,m_K^2 - 5/16\,m_\pi^2\,s + 5/12\,m_\pi^4 - 55/96
         \,m_K^2\,s + 5/12\,m_K^4
\nonumber\\&&
~~ + 5/32\,s^2 )
%\nonumber\\&&
       + P_{K\pi}\,
%\overline{B}(m_\pi^2,m_K^2,s) \,
 (  - 25/32\,m_\pi^2\,s^2 + 5/3\,m_\pi^4\,s - 5/
         6\,m_\pi^6 ))
\nonumber\\&&
       + \overline{B}(m_\eta^2,m_K^2,s) \, (  - 5/72\,m_\pi^2\,m_K^2 + 5/12\,m_\pi^2\,s - 5/18\,m_\pi^4 + 5/32\,
         m_K^2\,s + 5/72\,m_K^4 
\nonumber\\&&
~~ - 5/32\,s^2 )
%\nonumber\\&&
       + P_{K\pi}\,
%\overline{B}(m_\eta^2,m_K^2,s) \,
 (  - 5/32\,m_\pi^2\,s^2 + 5/12\,m_\pi^4\,s - 5/
         18\,m_\pi^6 ))
\nonumber\\&&
       + \overline{B_1}(m_\pi^2,m_K^2,s) \, ( 15/16\,m_\pi^2\,m_K^2 - 5/24\,m_\pi^4 + 25/48\,m_K^4 )
\nonumber\\&&
       + \overline{B_1}(m_\eta^2,m_K^2,s) \, (  - 5/16\,m_\pi^2\,m_K^2 - 5/48\,m_K^4 ).
\ea
\ba
\lefteqn{   M^{27}_5(s) =}&&
\nonumber\\&&
       + D_2^r \, (( 10/3\,m_\pi^2 )
%\nonumber\\&&
       + P_{K\pi}\,%D_2^r \,
 ( 10/3\,m_\pi^4 ))
%\nonumber\\&&
       + D_4^r \, ((  - 10\,m_\pi^2 - 5/2\,m_K^2 )
%\nonumber\\&&
       + P_{K\pi}\,%D_4^r \,
 (  - 10/3\,m_\pi^4 ))
\nonumber\\&&
       + D_5^r \, ( 20/3\,m_\pi^2 + 5/2\,m_K^2 )
%\nonumber\\&&
       + D_7^r \, ((  - 25/3\,m_\pi^2 - 10\,m_K^2 )
%\nonumber\\&&
       + P_{K\pi}\,%D_7^r \,
 (  - 5\,m_\pi^4 ))
\nonumber\\&&
       + D_{28}^r \, ( 10/3\,s )
%\nonumber\\&&
       + D_{29}^r \, (  - 5/12\,s )
%\nonumber\\&&
       + D_{30}^r \, ( 5/12\,s )
%\nonumber\\&&
       + P_{K\pi}\,(L_3^r-2L_5^r) \, ( 10/3\,m_\pi^2\,s )
\nonumber\\&&
       + ({1}/{(16\pi^2)}) \, (( 280/81\,m_\pi^2 + 1945/648\,m_K^2 - 55/54\,s )
%\nonumber\\&&
       + P_{K\pi}\,%({1}{(16\pi^2)} \,
 (  - 5/24\,m_\pi^2\,s + 5/4\,m_\pi^4 ))
\nonumber\\&&
       + \overline{A}(m_\pi^2) \, ((  - 1105/72 )
%\nonumber\\&&
       + P_{K\pi}\,%\overline{A}(m_\pi^2) \,
 (  - 95/24\,m_\pi^2 )
%\nonumber\\&&
       + P_{K\pi}^2\,%\overline{A}(m_\pi^2) \,
 ( 5/6\,m_\pi^4 ))
\nonumber\\&&
       + \overline{A}(m_K^2) \, ((  - 335/36 )
%\nonumber\\&&
       + P_{K\pi}\,%\overline{A}(m_K^2) \,
 (  - 85/24\,m_\pi^2 )
%\nonumber\\&&
       + P_{K\pi}^2\,%\overline{A}(m_K^2) \,
 (  - 5/6\,m_\pi^4 ))
\nonumber\\&&
       + \overline{A}(m_\eta^2) \, ((  - 35/24 )
%\nonumber\\&&
       + P_{K\pi}\,%\overline{A}(m_\eta^2) \,
 (  - 5/12\,m_\pi^2 ))
\nonumber\\&&
       + \overline{B}(m_\pi^2,m_\pi^2,s) \, (  - 5/6\,m_\pi^2 + 5/24\,s )
%\nonumber\\&&
       + P_{K\pi}\,%\overline{B}(m_\pi^2,m_\pi^2,s) \,
 ( 5/72\,m_\pi^2\,s - 5/18\,m_\pi^4 ))
\nonumber\\&&
       + \overline{B}(m_\pi^2,m_K^2,s) \, (  - 67/144\,m_\pi^2 - 43/288\,m_K^2 + 43/288\,s 
%\nonumber\\&&
       + P_{K\pi}\,%\overline{B}(m_\pi^2,m_K^2,s) \,
 ( 5/288\,m_\pi^2\,s - 5/72\,m_\pi^4 ))
\nonumber\\&&
       + \overline{B}(m_K^2,m_K^2,s) \, ((  - 5/18\,m_\pi^2 - 7/9\,m_K^2 + 7/36\,s )
%\nonumber\\&&
       + P_{K\pi}\,%\overline{B}(m_K^2,m_K^2,s) \,
 ( 5/72\,m_\pi^2\,s - 5/18\,m_\pi^4 ))
\nonumber\\&&
       + \overline{B}(m_\eta^2,m_K^2,s) \,( ( 7/24\,m_\pi^2 - 5/32\,m_K^2 + 1/32\,s )
%\nonumber\\&&
       + P_{K\pi}\,%\overline{B}(m_\eta^2,m_K^2,s) \,
 (  - 5/96\,m_\pi^2\,s + 5/24\,m_\pi^4 ))
\nonumber\\&&
       + \overline{B_1}(m_\pi^2,m_K^2,s) \, ( 19/72\,m_\pi^2 - 43/144\,m_K^2 )
%\nonumber\\&&
       + \overline{B_1}(m_\eta^2,m_K^2,s) \, (  - 1/18\,m_\pi^2 + 1/48\,m_K^2 ).
\nonumber\\
\ea
\ba
\lefteqn{   M^{27}_9(s) =}&&
\nonumber\\&&
       + \overline{B}(m_\pi^2,m_\pi^2,s) \, ((  - 19/18\,m_\pi^2 + 19/72\,s )
%\nonumber\\&&
       + P_{K\pi}\,%\overline{B}(m_\pi^2,m_\pi^2,s) \,
 ( 5/24\,m_\pi^2\,s - 5/6\,m_\pi^4 ))
\nonumber\\&&
       + \overline{B}(m_\pi^2,m_K^2,s) \,(   - 67/144\,m_\pi^2 - 43/288\,m_K^2 + 43/288\,s 
%\nonumber\\&&
       + P_{K\pi}\,%\overline{B}(m_\pi^2,m_K^2,s) \,
 ( 5/288\,m_\pi^2\,s - 5/72\,m_\pi^4 ))
\nonumber\\&&
       + \overline{B}(m_K^2,m_K^2,s) \,( (  - 5/9\,m_\pi^2 - 5/9\,m_K^2 + 5/36\,s )
%\nonumber\\&&
       + P_{K\pi}\,%\overline{B}(m_K^2,m_K^2,s) \,
 ( 5/36\,m_\pi^2\,s - 5/9\,m_\pi^4 ))
\nonumber\\&&
       + \overline{B}(m_\eta^2,m_K^2,s) \,( ( 7/24\,m_\pi^2 - 5/32\,m_K^2 + 1/32\,s )
%\nonumber\\&&
       + P_{K\pi}\,%\overline{B}(m_\eta^2,m_K^2,s) \,
 (  - 5/96\,m_\pi^2\,s + 5/24\,m_\pi^4 ))
\nonumber\\&&
       + \overline{B_1}(m_\pi^2,m_K^2,s) \, ( 19/72\,m_\pi^2 - 43/144\,m_K^2 )
%\nonumber\\&&
       + \overline{B_1}(m_\eta^2,m_K^2,s) \, (  - 1/18\,m_\pi^2 + 1/48\,m_K^2 ).
\nonumber\\
\ea
\ba
\lefteqn{   M^{27}_{10}(s) =}&&
\nonumber\\&&
       (D_1^r+2 D_2^r)  ( 8/3\,m_\pi^2\,m_K^2 )
%\nonumber\\&&
       + D_4^r  ( 14/3\,m_\pi^2 m_K^2 - 47/3\,m_\pi^2 s + 47/3\,m_\pi^4 - 13/3\,m_K^2\,s + m_K^4
          )
\nonumber\\&&
       + D_5^r \, (  - 10\,m_\pi^2\,m_K^2 + 47/3\,m_\pi^2\,s - 47/3\,m_\pi^4 + 13/3\,m_K^2\,s - 
         m_K^4 )
\nonumber\\&&
       + D_6^r \, (  - 2/3\,m_\pi^2\,m_K^2 + 1/3\,m_\pi^2\,s - 1/3\,m_\pi^4 - m_K^2\,s + m_K^4 )
\nonumber\\&&
       + D_7^r \, ( 50/3\,m_\pi^2\,m_K^2 - 34/3\,m_\pi^2\,s + 34/3\,m_\pi^4 - 52/3\,m_K^2\,s + 4\,
         m_K^4 )
\nonumber\\&&
       + D_{26}^r \, (  - 2/3\,m_\pi^2\,m_K^2 + m_\pi^2\,s - 5/3\,m_\pi^4 + 1/3\,m_K^2\,s - 1/3\,m_K^4
          )
\nonumber\\&&
       + D_{27}^r \, ( 4/3\,m_\pi^2\,m_K^2 - 2\,m_\pi^2\,s - 2/3\,m_\pi^4 - 2/3\,m_K^2\,s - 2/3\,m_K^4
          + 4/3\,s^2 )
\nonumber\\&&
       + D_{28}^r \, (  - 12\,m_\pi^2\,m_K^2 + 18\,m_\pi^2\,s - 18\,m_\pi^4 + 6\,m_K^2\,s - 2\,m_K^4 - 
         4\,s^2 )
\nonumber\\&&
       + D_{29}^r \, ( 8/3\,m_\pi^2\,m_K^2 - 4\,m_\pi^2\,s + 8/3\,m_\pi^4 - 4/3\,m_K^2\,s + 4/3\,s^2 )
\nonumber\\&&
       + D_{30}^r \, (  - 8/3\,m_\pi^2\,m_K^2 + 4\,m_\pi^2\,s - 20/3\,m_\pi^4 + 4/3\,m_K^2\,s - 4/3\,
         m_K^4 )
\nonumber\\&&
       + D_{31}^r \, (  - 8/3\,m_\pi^2\,m_K^2 + 4\,m_\pi^2\,s - 20/3\,m_\pi^4 + 4/3\,m_K^2\,s - 4/3\,
         m_K^4 )
\nonumber\\&&
       + L_1^r \, ( 64/3\,m_\pi^2\,m_K^2 - 32\,m_\pi^2\,s + 160/3\,m_\pi^4 - 32/3\,m_K^2\,s + 32/3\,
         m_K^4 )
\nonumber\\&&
       + L_2^r \, (  - 32/3\,m_\pi^2\,m_K^2 + 16\,m_\pi^2\,s + 16/3\,m_\pi^4 + 16/3\,m_K^2\,s + 16/3
         \,m_K^4 - 32/3\,s^2 )
\nonumber\\&&
       + L_3^r \, ( 32/3\,m_\pi^2\,m_K^2 - 16\,m_\pi^2\,s + 80/3\,m_\pi^4 - 16/3\,m_K^2\,s + 16/3\,
         m_K^4 )
\nonumber\\&&
       + L_4^r \, ( 32/3\,m_\pi^2\,m_K^2 + 32/3\,m_\pi^2\,s - 32/3\,m_\pi^4 )
%\nonumber\\&&
       + L_5^r \, ( 32/3\,m_\pi^2\,s - 32/3\,m_\pi^4 )
\nonumber\\&&
       + ({1}/{(16\pi^2)}) \, (  - 643/162\,m_\pi^2\,m_K^2 - 49/324\,m_\pi^2\,s + 31/108\,m_\pi^4 + 1273/324
         \,m_K^2\,s
\nonumber\\&&
~~ - 439/324\,m_K^4 + 34/27\,s^2 )
\nonumber\\&&
       + \overline{A}(m_\pi^2) \,( ( 2075/72\,m_\pi^2 + 493/72\,m_K^2 - 329/12\,s )
%\nonumber\\&&
       + P_{K\pi}\,%\overline{A}(m_\pi^2) \,
 (  - 35/24\,m_\pi^2\,s + 25/9\,m_\pi^4 ))
\nonumber\\&&
       + \overline{A}(m_K^2) \, (( 575/36\,m_\pi^2 + 149/36\,m_K^2 - 49/3\,s )
%\nonumber\\&&
       + P_{K\pi}\,%\overline{A}(m_K^2) \,
 ( 5/6\,m_\pi^2\,s - 10/9\,m_\pi^4 ))
\nonumber\\&&
       + \overline{A}(m_\eta^2) \, (( 29/24\,m_\pi^2 + 19/24\,m_K^2 - 9/4\,s )
%\nonumber\\&&
       + P_{K\pi}\,%\overline{A}(m_\eta^2) \,
 ( 5/8\,m_\pi^2\,s - 5/3\,m_\pi^4 ))
\nonumber\\&&
       + \overline{B}(m_\pi^2,m_\pi^2,s) \, ( m_\pi^2\,m_K^2 - 13/2\,m_\pi^2\,s + 13/3\,m_\pi^4 - 1/2\,m_K^2\,s + 
         13/6\,s^2 )
\nonumber\\&&
       + \overline{B}(m_\pi^2,m_K^2,s) \, ( 8/3\,m_\pi^2\,m_K^2 - 13/3\,m_\pi^2\,s + 13/6\,m_\pi^4 - 8/3\,m_K^2\,
         s + 1/2\,m_K^4 + 13/6\,s^2 ).
\nonumber\\&&
\ea
\ba
\lefteqn{   M^{27}_{11}(s) =}&&
\nonumber\\&&
       + (D_{26}^r+2 D_{27}^r+6 D_{28}^r+4 D_{30}^r+4 D_{31}^r) \, ( 1/3\,s^2 )
%\nonumber\\&&
       + (2 L_1^r+ L_2^r+ L_3^r) \, (  - 16/3\,s^2 )
\nonumber\\&&
       + ({1}/{(16\pi^2)}) \, (  - 11/162\,m_\pi^2\,s + 11/162\,m_K^2\,s - 17/27\,s^2 )
\nonumber\\&&
       + \overline{B}(m_\pi^2,m_\pi^2,s) \,( (  - 5/12\,m_\pi^2\,m_K^2 + 5/2\,m_\pi^2\,s - 2\,m_\pi^4 - 1/6\,m_K^2\,
         s - 1/2\,s^2 )
\nonumber\\&&
~~       + P_{K\pi}\,%\overline{B}(m_\pi^2,m_\pi^2,s) \,
 (  - 5/4\,m_\pi^2\,s^2 + 35/12\,m_\pi^4\,s - 5/
         3\,m_\pi^6 ))
\nonumber\\&&
       + \overline{B}(m_\pi^2,m_K^2,s) \,( ( 7/24\,m_\pi^2\,m_K^2 - 13/24\,m_\pi^2\,s + 7/12\,m_\pi^4 + 19/48\,
         m_K^2\,s - 3/8\,m_K^4
\nonumber\\&&
~~ - 1/48\,s^2 )
      + P_{K\pi}\,%\overline{B}(m_\pi^2,m_K^2,s) \,
 ( 15/16\,m_\pi^2\,s^2 - 5/2\,m_\pi^4\,s + 5/3\,
         m_\pi^6 ))
\nonumber\\&&
       + \overline{B}(m_K^2,m_K^2,s) \, (  - 3/8\,m_\pi^2\,s - 3/8\,m_K^2\,s + 3/8\,s^2 )
\nonumber\\&&
       + \overline{B}(m_\eta^2,m_K^2,s) \,( (  - 5/72\,m_\pi^2\,m_K^2 - 5/6\,m_\pi^2\,s + 5/9\,m_\pi^4 - 3/16\,
         m_K^2\,s - 7/72\,m_K^4 
\nonumber\\&&
~~ + 5/16\,s^2 )
       + P_{K\pi}\,%\overline{B}(m_\eta^2,m_K^2,s) \,
 ( 5/16\,m_\pi^2\,s^2 - 5/6\,m_\pi^4\,s + 5/9\,
         m_\pi^6 ))
\nonumber\\&&
       + \overline{B}(m_\eta^2,m_\eta^2,s) \,( (  - 13/36\,m_\pi^2\,m_K^2 + 5/12\,m_\pi^2\,s - 5/9\,m_\pi^4 )
%\nonumber\\&&
       + P_{K\pi}\,%\overline{B}(m_\eta^2,m_\eta^2,s) \,
 ( 5/12\,m_\pi^4\,s - 5/9\,m_\pi^6 ))
\nonumber\\&&
       + \overline{B_1}(m_\pi^2,m_K^2,s) \, (  - 13/24\,m_\pi^2\,m_K^2 + 1/12\,m_\pi^4 - 3/8\,m_K^4 )
\nonumber\\&&
       + \overline{B_1}(m_\eta^2,m_K^2,s) \, ( 7/8\,m_\pi^2\,m_K^2 - 1/24\,m_K^4 ).
\ea
\ba
\lefteqn{   M^{27}_{12}(s) =}&&
\nonumber\\&&
       + \overline{B}(m_\pi^2,m_\pi^2,s) \, (  - 13/9\,m_\pi^2 + 13/36\,s )
\nonumber\\&&
       + \overline{B}(m_\pi^2,m_K^2,s) \,( (  - 8/9\,m_\pi^2 - 41/144\,m_K^2 + 41/144\,s )
%\nonumber\\&&
       + P_{K\pi}\,
%\overline{B}(m_\pi^2,m_K^2,s) \,
 ( 5/144\,m_\pi^2\,s - 5/36\,m_\pi^4 ))
\nonumber\\&&
       + \overline{B}(m_K^2,m_K^2,s) \,( ( 5/9\,m_\pi^2 - 5/18\,m_K^2 + 5/72\,s )
%\nonumber\\&&
       + P_{K\pi}\,
%\overline{B}(m_K^2,m_K^2,s) \, 
(  - 5/36\,m_\pi^2\,s + 5/9\,m_\pi^4 ))
\nonumber\\&&
       + \overline{B}(m_\eta^2,m_K^2,s) \,( (  - 7/24\,m_\pi^2 - 55/48\,m_K^2 + 11/48\,s )
%\nonumber\\&&
       + P_{K\pi}\,
%\overline{B}(m_\eta^2,m_K^2,s) \,
 ( 5/48\,m_\pi^2\,s - 5/12\,m_\pi^4 ))
\nonumber\\&&
       + \overline{B_1}(m_\pi^2,m_K^2,s) \, ( 1/2\,m_\pi^2 - 41/72\,m_K^2 )
%\nonumber\\&&
       + \overline{B_1}(m_\eta^2,m_K^2,s) \, (  - 1/12\,m_\pi^2 + 11/72\,m_K^2 ).
\nonumber\\&&
\ea


\begin{thebibliography}{99}

\bibitem{Weinberg}
S.~Weinberg,
%``Phenomenological Lagrangians,''
PhysicaA {\bf 96} (1979) 327.
%%CITATION = PHYSA,A96,327;%%

\bibitem{GL1}
J.~Gasser and H.~Leutwyler,
%``Chiral Perturbation Theory To One Loop,''
Annals Phys.\  {\bf 158} (1984) 142.
%%CITATION = APNYA,158,142;%%

\bibitem{GL2}
J.~Gasser and H.~Leutwyler,
%``Chiral Perturbation Theory: Expansions In The Mass Of The Strange Quark,''
Nucl.\ Phys.\ B {\bf 250} (1985) 465.
%%CITATION = NUPHA,B250,465;%%

\bibitem{chptlectures}
Pich, A., Lectures at Les Houches Summer School in
Theoretical Physics, Session 68: Probing the Standard Model of Particle
Interactions, Les Houches, France, 28 Jul - 5 Sep 1997,
[hep-ph/9806303];\\
%%CITATION = HEP-PH 9806303;%%
Ecker, G.,
Lectures given at Advanced School on Quantum Chromodynamics (QCD 2000),
Benasque, Huesca, Spain, 3-6 Jul 2000,
[hep-ph/0011026].
%%CITATION = HEP-PH 0011026;%%

\bibitem{KMW1}
J.~Kambor, J.~Missimer and D.~Wyler,
%``The Chiral Loop Expansion Of The Nonleptonic Weak Interactions Of Mesons,''
Nucl.\ Phys.\ B {\bf 346} (1990) 17.
%%CITATION = NUPHA,B346,17;%%

\bibitem{KMW2}
J.~Kambor, J.~Missimer and D.~Wyler,
%``K $\to$ 2 Pi And K $\to$ 3 Pi Decays In Next-To-Leading Order Chiral Perturbation Theory,''
Phys.\ Lett.\ B {\bf 261} (1991) 496.
%%CITATION = PHLTA,B261,496;%%

\bibitem{KDHMW}
J.~Kambor, J.~F.~Donoghue, B.~R.~Holstein, J.~Missimer and D.~Wyler,
%``Chiral Symmetry Tests In Nonleptonic K Decay,''
Phys.\ Rev.\ Lett.\  {\bf 68} (1992) 1818.
%%CITATION = PRLTA,68,1818;%%

\bibitem{chptweakreviews}
G.~Ecker,
%``Chiral perturbation theory,''
Prog.\ Part.\ Nucl.\ Phys.\  {\bf 35} (1995) 1
[hep-ph/9501357];\\
%%CITATION = HEP-PH 9501357;%%
A.~Pich,
%``Chiral perturbation theory,''
Rept.\ Prog.\ Phys.\  {\bf 58} (1995) 563
[hep-ph/9502366];\\
%%CITATION = HEP-PH 9502366;%%
E.~de Rafael,
%``Chiral Lagrangians and kaon CP violation,''
Lectures given at Theoretical Advanced Study Institute in Elementary
Particle Physics (TASI 94): CP Violation and the limits of the Standard Model,
Boulder, CO, 29 May - 24 Jun 1994,
hep-ph/9502254.
%%CITATION = HEP-PH 9502254;%%

\bibitem{Cronin}
J.~A.~Cronin,
%``Phenomenological Model Of Strong And Weak Interactions In Chiral U(3) X U(3),''
Phys.\ Rev.\  {\bf 161} (1967) 1483.
%%CITATION = PHRVA,161,1483;%%

\bibitem{k3piold}
B.~R.~Holstein,
%``Current-Algebra Model Of Delta-I=3/2 Effects In Nonleptonic Kaon Decay,''
Phys.\ Rev.\  {\bf 183} (1969) 1228;\\
%%CITATION = PHRVA,183,1228;%%
J.~F.~Donoghue, E.~Golowich and B.~R.~Holstein,
%``Kaon Decays And A Determination Of The Scale Of Chiral Symmetry,''
Phys.\ Rev.\ D {\bf 30} (1984) 587;\\
%%CITATION = PHRVA,D30,587;%%
H.~Y.~Cheng, C.~Y.~Cheung and W.~B.~Yeung,
%``Toward The Understanding Of K $\to$ 3 Pi Decays In Chiral Perturbation Theory,''
Mod.\ Phys.\ Lett.\ A {\bf 4} (1989) 869;
%%CITATION = MPLAE,A4,869;%%
%\bibitem{Cheng:1988ps}
%H.~Y.~Cheng, C.~Y.~Cheung and W.~B.~Yeung,
%``Analysis Of K $\to$ 3 Pi Decays In Chiral Perturbation Theory,''
Z.\ Phys.\ C {\bf 43} (1989) 391;\\
%%CITATION = ZEPYA,C43,391;%%
S.~Fajfer and J.~M.~Gerard,
%``A Simple Chiral Lagrangian Approach To K $\to$ Pi Pi Pi Decays And Epsilon-Prime+0-,''
Z.\ Phys.\ C {\bf 42} (1989) 425.
%%CITATION = ZEPYA,C42,425;%%

\bibitem{knechtetal}
M.~Knecht, B.~Moussallam, J.~Stern and N.~H.~Fuchs,
%``The Low-energy pi pi amplitude to one and two loops,''
Nucl.\ Phys.\ B {\bf 457} (1995) 513
[hep-ph/9507319].
%%CITATION = HEP-PH 9507319;%%

\bibitem{Esposito}
G.~Esposito-Farese,
%``Right Invariant Metrics On SU(3) And One Loop Divergences In Chiral Perturbation Theory,''
Z.\ Phys.\ C {\bf 50} (1991) 255.
%%CITATION = ZEPYA,C50,255;%%

\bibitem{EKW}
G.~Ecker, J.~Kambor and D.~Wyler,
%``Resonances in the weak chiral Lagrangian,''
Nucl.\ Phys.\ B {\bf 394} (1993) 101.
%%CITATION = NUPHA,B394,101;%%

\bibitem{BPP}
J.~Bijnens, E.~Pallante and J.~Prades,
%``Obtaining K $\to$ pi pi from off-shell K $\to$ pi amplitudes,''
Nucl.\ Phys.\ B {\bf 521} (1998) 305
[hep-ph/9801326].
%%CITATION = HEP-PH 9801326;%%

\bibitem{Zemach}
C. Zemach,
Phys.Rev. {\bf 133} (1964) 1201.
%%CITATION = PHRVA,133,1201;%%

\bibitem{Weinberg2}
S. Weinberg,
Phys. Rev. Lett. {\bf 17} (1966) 61.
%%CITATION = PRLTA,17,61;%%

\bibitem{DAmbrosio}
G.~D'Ambrosio, G.~Isidori, A.~Pugliese and N.~Paver,
%``Strong rescattering in K $\to$ 3 pi decays and low-energy meson dynamics,''
Phys.\ Rev.\ D {\bf 50} (1994) 5767
[Erratum-ibid.\ D {\bf 51} (1994) 3975]
[hep-ph/9403235].
%%CITATION = HEP-PH 9403235;%%

\bibitem{Devlin}
T.~J.~Devlin and J.~O.~Dickey,
%``Weak Hadronic Decays: K $\to$ 2 Pi And K $\to$ 3 Pi,''
Rev.\ Mod.\ Phys.\  {\bf 51} (1979) 237.
%%CITATION = RMPHA,51,237;%%

\bibitem{Cheshkov}
C.~Cheshkov,
%``Nonleptonic kaon decays: Theory vs. experiment,''
hep-ph/0105131.
%%CITATION = HEP-PH 0105131;%%

\bibitem{PDG2000}
D.~E.~Groom {\it et al.}  [Particle Data Group Collaboration],
%``Review Of Particle Physics,''
Eur.\ Phys.\ J.\ C {\bf 15} (2000) 1.
%%CITATION = EPHJA,C15,1;%%

\bibitem{CPLEAR}
R.~Adler {\it et al.}  [CPLEAR Collaboration],
%``CPLEAR results on the CP parameters of neutral kaons decaying to  pi+ pi- pi0,''
Phys.\ Lett.\ B {\bf 407} (1997) 193;\\
%%CITATION = PHLTA,B407,193;%%
R.~Adler {\it et al.}  [CPLEAR Collaboration],
%``Observation of the CP conserving K(S) $\to$ pi+ pi- pi0 decay amplitude,''
Phys.\ Lett.\ B {\bf 374} (1996) 313;\\
%%CITATION = PHLTA,B374,313;%%
A.~Angelopoulos {\it et al.}  [CPLEAR Collaboration],
%``The neutral kaon decays to pi+ pi- pi0: A detailed analysis of the  CPLEAR data,''
Eur.\ Phys.\ J.\ C {\bf 5} (1998) 389.
%%CITATION = EPHJA,C5,389;%%

\bibitem{E731}
S.~Somalwar {\it et al.}  [Fermilab E-731 Collaboration],
%``A Measurement Of The Quadratic Slope Parameter In The K(L) $\to$ 3 Pi0 Decay Dalitz Plot,''
Phys.\ Rev.\ Lett.\  {\bf 68} (1992) 2580.
%%CITATION = PRLTA,68,2580;%%

\bibitem{NA48}
A.~Lai {\it et al.}  [NA48 Collaboration],
%``Measurement of the quadratic slope parameter in the K(L) $\to$ 3pi0 decay  Dalitz plot,''
Phys.\ Lett.\ B {\bf 515} (2001) 261
[hep-ex/0106075].
%%CITATION = HEP-EX 0106075;%%

\bibitem{Zou}
G.~B.~Thomson {\it et al.},
%``Measurement Of The Amplitude Of The CP Conserving Decay K0(S) $\to$ Pi+ Pi- Pi0,''
Phys.\ Lett.\ B {\bf 337} (1994) 411;\\
%%CITATION = PHLTA,B337,411;%%
Y.~Zou {\it et al.},
%``New Measurement Of The Amplitude Of The CP Conserving Decay K0(S) $\to$ Pi+ Pi- Pi0,''
Phys.\ Lett.\ B {\bf 369} (1996) 362.
%%CITATION = PHLTA,B369,362;%%

\bibitem{Batusov}
V.~Y.~Batusov {\it et al.},
%``Measurement Of The Dalitz Plot Slope Parameters For K+ $\to$ Pi+ Pi0 Pi0  Decay,''
Nucl.\ Phys.\ B {\bf 516} (1998) 3.
%%CITATION = NUPHA,B516,3;%%

\bibitem{Bolotov}
V.~N.~Bolotov {\it et al.},
%``Study Of The Decay K- $\to$ Pi- Pi0 Pi0. (In Russian),''
Sov.\ J.\ Nucl.\ Phys.\  {\bf 44} (1986) 73
[Yad.\ Fiz.\  {\bf 44} (1986) 117].
%%CITATION = SJNCA,44,73;%%

\bibitem{ABT2}
G.~Amoros, J.~Bijnens and P.~Talavera,
%``Low energy constants from K(l4) form-factors,''
Phys.\ Lett.\ B {\bf 480} (2000) 71
[hep-ph/9912398];
%%CITATION = HEP-PH 9912398;%%
%G.~Amoros, J.~Bijnens and P.~Talavera,
%``K(l4) form-factors and pi pi scattering,''
Nucl.\ Phys.\ B {\bf 585} (2000) 293
[Erratum-ibid.\ B {\bf 598} (2001) 665]
[hep-ph/0003258];
%%CITATION = HEP-PH 0003258;%%
%G.~Amoros, J.~Bijnens and P.~Talavera,
%``QCD isospin breaking in meson masses, decay constants and quark mass  ratios,''
Nucl.\ Phys.\ B {\bf 602} (2001) 87
[hep-ph/0101127].
%%CITATION = HEP-PH 0101127;%%

\bibitem{BG}
J.~Bijnens and J.~Gasser,
``Eta decays at and beyond $p^4$ in chiral perturbation theory,''
[hep-ph/0202242], proceedings of the
Workshop on Eta Physics: Prospects of Precision Measurements with the
CELSIUS/WASA Facility, Uppsala, Sweden, 25-27 Oct 2001.
%%CITATION = HEP-PH 0202242;%%

% isospin breaking in K to 2pi
\bibitem{iso}
G.~Ecker, G.~Isidori, G.~Muller, H.~Neufeld and A.~Pich,
%``Electromagnetism in nonleptonic weak interactions,''
Nucl.\ Phys.\ B {\bf 591} (2000) 419
[hep-ph/0006172];\\
%%CITATION = HEP-PH 0006172;%%
V.~Cirigliano, J.~F.~Donoghue and E.~Golowich,
%``Electromagnetic corrections to K $\to$ pi pi. II: Dispersive matching,''
Phys.\ Rev.\ D {\bf 61} (2000) 093002
[hep-ph/9909473];
%%CITATION = HEP-PH 9909473;%%
%V.~Cirigliano, J.~F.~Donoghue and E.~Golowich,
%``Electromagnetic corrections to K $\to$ pi pi. I: Chiral perturbation  theory,''
Phys.\ Rev.\ D {\bf 61} (2000) 093001
[Erratum-ibid.\ D {\bf 63} (2000) 059903]
[hep-ph/9907341];\\
%%CITATION = HEP-PH 9907341;%%
S.~Gardner and G.~Valencia,
%``Additional isospin-breaking effects in $\epsilon^\prime/\epsilon$,''
Phys.\ Lett.\ B {\bf 466} (1999) 355
[hep-ph/9909202];\\
%%CITATION = HEP-PH 9909202;%%
C.~E.~Wolfe and K.~Maltman,
%``Strong isospin-breaking effects on the Delta(I) = 3/2 amplitude in  K $\to$ 2pi at next-to-leading order in the chiral expansion,''
Phys.\ Lett.\ B {\bf 482} (2000) 77
[hep-ph/9912254];\\
%%CITATION = HEP-PH 9912254;%%
G.~Ecker, G.~Muller, H.~Neufeld and A.~Pich,
%``pi0 eta mixing and CP violation,''
Phys.\ Lett.\ B {\bf 477} (2000) 88
[hep-ph/9912264].
%%CITATION = HEP-PH 9912264;%%

\bibitem{ABT1}
G.~Amoros, J.~Bijnens and P.~Talavera,
%``Two-point functions at two loops in three flavour chiral perturbation  theory,''
Nucl.\ Phys.\ B {\bf 568} (2000) 319
[hep-ph/9907264].
%%CITATION = HEP-PH 9907264;%%

\end{thebibliography}
\end{document}